\documentclass[sigplan,screen,nonacm]{acmart}
\renewcommand\footnotetextcopyrightpermission[1]{}
\usepackage{tikz}
\usepackage{amsmath}
\usepackage{filecontents}
\usepackage{subcaption}

\usepackage{multirow} 
\usepackage{booktabs}

\usepackage{algorithm}  

\usepackage{amsmath,amsfonts}
\usepackage{bm}
\usepackage{algorithmicx}  
\usepackage{amsmath}
\usepackage{algpseudocode} 
\usepackage[T1]{fontenc}
\usepackage{hyperref}
\usepackage{comment}
\usepackage{mathrsfs}
\usepackage{enumitem}

\usepackage{listings}
\newcommand{\proj}[1]{$Vortex$}

\newcommand{\abst}[1]{\textit{rKernel}}

\renewcommand{\paragraph}[1]{\vspace*{.1cm}\noindent\textbf{#1}\hspace*{.1cm}}

\begin{document}

\title{Vortex: Efficient Sample-Free Dynamic Tensor Program Optimization via Hardware-aware \\ Strategy Space Hierarchization
}

\date{}

\author{Yangjie Zhou\textsuperscript{*}}
\email{jasonyjzhou@tencent.com}
\affiliation{
    \institution{Tencent}
    \country{}
}

\author{Honglin Zhu\textsuperscript{*}}
\email{honglinzhu@tencent.com}
\affiliation{
    \institution{Tencent}
    \country{}
}

\author{Qian Qiu}
\email{qianqiu@tencent.com}
\affiliation{
    \institution{Tencent}
    \country{}
}

\author{Weihao Cui}
\email{weihao@sjtu.edu.cn}
\affiliation{
    \institution{Shanghai Jiao Tong University}
    \country{}
}

\author{Zihan Liu}
\email{altair.liu@sjtu.edu.cn}
\affiliation{
    \institution{Shanghai Jiao Tong University}
    \country{}
}

\author{Cong Guo}
\email{guocong@sjtu.edu.cn}
\affiliation{
    \institution{Shanghai Jiao Tong University}
    \country{}
}

\author{Siyuan Feng}
\email{hzfengsy@sjtu.edu.cn}
\affiliation{
    \institution{Shanghai Jiao Tong University}
    \country{}
}

\author{Jintao Meng}
\email{jt.meng@siat.ac.cn}
\affiliation{
    \institution{Shenzhen Institute of Advanced Technology}
    \country{}
}

\author{Haidong Lan}
\email{haidonglan@taichi.graphics}
\affiliation{
    \institution{Taichi Graphics}
    \country{}
}

\author{Jingwen Leng}
\email{leng-jw@cs.sjtu.edu.cn}
\affiliation{
    \institution{Shanghai Jiao Tong University}
    \country{}
}

\author{Wenxi Zhu\textsuperscript{\dag}}
\email{wenxizhu@tencent.com}
\affiliation{
    \institution{Tencent}
    \country{}
}

\author{Minwen Deng\textsuperscript{\dag}}
\email{danierdeng@tencent.com}
\affiliation{
    \institution{Tencent}
    \country{}
}

\thanks{\textsuperscript{*}Yangjie Zhou and Honglin Zhu contributed equally to this work.}
\thanks{\textsuperscript{\dag}Wenxi Zhu and Minwen Deng are the corresponding auhtors of this paper.}

\begin{abstract}

Dynamic-shape deep neural networks (DNNs) are rapidly evolving, attracting attention for their ability to handle variable input sizes in real-time applications.
However, existing compilation optimization methods for such networks often rely heavily on predefined samples to guide the compilation process, which restricts their adaptability and efficiency. These sample-driven methods struggle to efficiently manage the diverse and unpredictable shapes encountered in real-world scenarios, often resulting in suboptimal performance.

To tackle these issues, we introduce \proj{}, a hardware-driven and sample-free compiler tailored for dynamic-shape tensor programs.
\proj{} capitalizes on detailed hardware information and hierarchizes the strategy space to facilitate high-performance code generation without relying on runtime shape samples.
It features a unique bidirectional compilation workflow, combining top-down abstraction for aligning tensor program execution with hardware hierarchies and bottom-up kernel construction to narrow the search space, enabling \proj{} to achieve remarkable efficiency.
Comprehensive evaluations confirm that \proj{} reduces compilation time by $176\times$ compared to the existing dynamic-shape compiler. Additionally, it substantially outperforms existing vendor-provided libraries and dynamic-shape compilers on both CPU and GPU platforms, delivering speedups of $2.53\times$ and $3.01\times$, respectively.

\end{abstract}

\thispagestyle{empty}
\pagestyle{plain}

\maketitle

\section{Introduction}

Efficient optimization of tensor programs is crucial in accelerating Deep Neural Network (DNN) models~\cite{lecun2015deep} and large language models (LLMs)~\cite{llm_survey}. 
Modern frameworks, e.g., PyTorch~\cite{pytorch} and TensorFlow~\cite{tensorflow}, and compilers such as TVM~\cite{tvm} significantly rely on tensor programs for DNN computational operator abstractions.
Traditional compilers~\cite{tvm,halide, ansor,tensorir} have primarily focused on optimizing \textbf{static-shape} DNNs, where tensor computations involve fixed-shape inputs and outputs at runtime. 
In contrast, the emergence of \textbf{dynamic-shape} DNNs, capable of handling variable input shapes during runtime, has become a significant area of interest~\cite{dynamic_survey,dietcode,bladedisc, nimble}. 
For example, large language models~\cite{llm_survey} based on Transformers~\cite{attention} often adopt variable sequence lengths, necessitating {dynamic-shape} tensor computation.
Effectively managing tensor computation with dynamic shapes is pivotal for optimizing neural network performance. 
However, the flexibility introduced by dynamic shapes poses challenges for optimizing tensor programs.

\begin{figure}[t]
    \centering
    \includegraphics[width=0.98\linewidth]{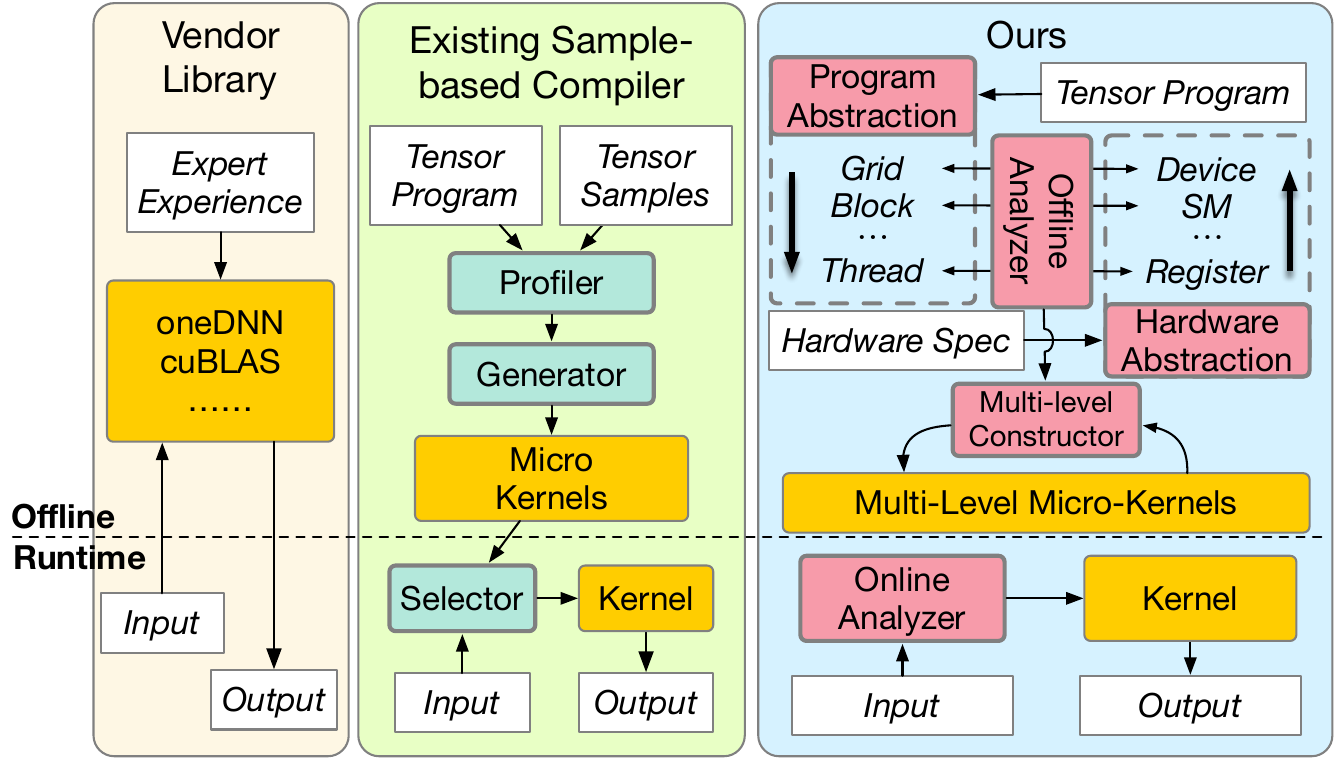}

        \caption{Comparison of \proj{} with existing methods.}

    \label{fig:intro}

\end{figure}

As illustrated in \autoref{fig:intro}, two main types of solutions have evolved to address the complexities of optimizing dynamic-shape tensor program computation.
The first solution is the vendor-provided library, exemplified by oneDNN~\cite{onednn} for Intel CPUs and cuBLAS~\cite{cublas} for Nvidia GPUs, offering handcrafted implementations of DNN operators.
While effective in certain scenarios, these libraries face limitations due to their empirical programming strategy, which does not offer the necessary flexibility for broad adaptability~\cite{dietcode}.
Additionally, the high development cost required to create these handcrafted solutions further constrains their efficiency~\cite{ansor}.

The second category encompasses existing \textbf{sample-driven} dynamic-shape~\cite{dietcode, nimble} compilers, as shown in \autoref{fig:intro} middle.
Similar in workflow to static-shape compilers~\cite{tvm, ansor}, these dynamic compilers utilize the tensor program and tensor samples as inputs to generate executable kernels.
Typically, tensor compilers construct a substantial search space to implement optimization strategies such as loop partitioning, fusion, and reordering~\cite{tvm,ansor}.
Existing dynamic-shape compilers~\cite{dietcode, nimble} usually adopt a shape-generic search space by using tensor samples to represent shape information and auto-tuning micro-kernels for each specific sample at the offline phase.
In the runtime, the dynamic compilers adopt a selector to integrate the micro-kernels for the computation process.
However, their reliance on predefined shape samples limits their flexibility and effectiveness, particularly when tensor shapes fall outside the predefined range. 
This limitation can result in performance degradations of up to $4\times$ for unsampled shapes (\S\ref{sec:mot:sample}).
Furthermore, these approaches necessitate frequent re-tuning through profiling on actual hardware to accommodate sample variations, which incurs considerable overhead, often taking hours to days~\cite{dietcode,nimble}.

Recognizing the limitations of existing methodologies, which predominantly rely on sample-based compilation~\cite{dietcode,nimble} or intensive manual implementation~\cite{cublas,cudnn}, we underscore the need for an innovative approach that fully exploits the capabilities of hardware architecture.
By adopting a hierarchical approach, we decouple the kernel into multiple levels: during the offline stage, we leverage detailed hardware information to construct hardware-friendly micro-kernels. Subsequently, during runtime, we utilize the shape information to select and configure micro-kernels dynamically, crafting shape-friendly implementations tailored to the needs of dynamic-shape tensor programs.
This methodology not only enhances tensor program execution but also eliminates the reliance on predefined shape samples, marking a significant shift towards dynamic, real-time compilation strategies.

In this work, we propose \proj{}, a \textbf{hardware-driven} and \textbf{sample-free} compiler tailored for dynamic-shape tensor programs.
To achieve these goals, we employ a novel \textbf{bidirectional} method to integrate hardware information into the compilation process.
Specifically, this bidirectional approach ensures a top-down alignment software tensor program with hardware hierarchy and a bottom-up construction that dynamically adapts to changing tensor shapes, as demonstrated in \autoref{fig:intro}.

First, \proj{} employs a \textbf{top-down abstraction} strategy to recursively decouple the tensor program, ensuring alignment with the hardware's hierarchical architecture.
For instance, in CUDA kernel on GPU, this process decouples the tensor program, mapping it to the software programming model’s Grid, Block, and Thread levels and the corresponding hardware structures: device, streaming multiprocessors (SMs), and registers~\cite{a100_whitepaper}.
By mirroring the hardware's hierarchy organization, \proj{} optimizes the execution implementation and resource allocation specific to each hardware level. This tailored approach ensures that the software exploits the full potential of the hardware architecture, leading to a significant boost in kernel performance.

Second, \proj{} can generate the multi-level micro-kernels using \textbf{bottom-up construction} strategy.
\proj{} uses hardware parameter information to prune a vast amount of strategy space, leading to more efficient code generation. 
An iterative constructor, progressing from lower to higher levels, utilizes this pruned strategy space to guide the construction of the kernels. This approach reduces the search space and accelerates the development process, thereby enhancing efficiency.
Furthermore, \proj{} incorporates a hybrid analyzer that incorporates analytical~\cite{chimera,micro23_path} and empirical approaches~\cite{tvm,ansor} to evaluate the performance of various strategies, thereby enabling high-performance code generation with minimal system overhead.

\proj{} stands out due to its superior runtime performance and significantly reduced compilation overhead.
We conduct a comprehensive evaluation of \proj{}, including a detailed comparative analysis with state-of-the-art vendor-provided libraries (oneDNN~\cite{onednn}, ONNX Runtime~\cite{onnxruntime}, cuBLAS~\cite{cublas}, cuDNN~\cite{cudnn} and CUTLASS~\cite{cutlass}) and dynamic-shape compiler DietCode~\cite{dietcode}. These assessments are conducted across various hardware platforms, including Intel CPUs~\cite{intel_xeon_whitepaper} and Nvidia GPUs~\cite{a100_whitepaper}, and at both operator and model levels. 
Notably, the performance of \proj{} showcases a remarkable speed, averaging $\mathbf{2.53\times}$ and $\mathbf{3.01\times}$ faster than vendor-provided libraries and DietCode, respectively. 
Additionally, \proj{} significantly accelerates the offline compilation, achieving $\mathbf{176\times}$ improvement over DietCode. 
Overall, the results highlight \proj{}'s ability to outperform existing solutions across both CPU and GPU platforms.

In general, our work makes the following contributions:
\begin{itemize}[itemsep=0pt, parsep=0pt, labelsep=5pt, leftmargin=*, topsep=0pt,partopsep=0pt]
    \item We identify hardware-driven opportunities as pivotal factors for optimizing dynamic-shape compilation, highlighting the limitations of existing sample-driven solutions in terms of flexibility and functionality.
    \item We propose \abst{}, a top-down unified abstraction to decouple tensor programs and align their strategy space with hardware hierarchies, ensuring that the software fully exploits the potential of the backend hardware.
    \item We introduce a bottom-up kernel construction approach, leveraging hardware information to prune the strategy space efficiently. 
    This method facilitates the generation of kernels for dynamic-shape tensor programs without reliance on runtime shape samples.

    \item We implement \proj{}, a novel dynamic-shape compiler. Our comprehensive evaluation demonstrates \proj{}'s superiority over existing dynamic-shape compilers and manual optimization techniques on CPU and GPU platforms.
\end{itemize}

\section{Background and Motivation}

\label{sec:background}

In this section, we begin by presenting a concise background on dynamic-shape tensor programs, outlining their definition and real-world computational scenarios. We then move to delineate the inherent constraints of the existing sample-driven optimization approach. Finally, we explore the opportunities and challenges in hardware-driven solutions.

\subsection{Dynamic-shape Tensor Program}

Tensor programs act as an operator-level abstraction, they are widely used across different tasks in neural network computation~\cite{pytorch,tensorflow,tvm,ansor}. 
Conventional tensor programs inherently incorporate static shape information as an integral part of their input. On the contrary, dynamic-shape tensor programs enable processing tensor programs with unknown shapes~\cite{dynamic_survey}. 
Dynamic-shape tensor programs have arisen in response to a twofold demand, driven by \textit{intrinsic data format dynamism} and \textit{system execution and scheduling strategies}. 

\begin{figure}[t]
    \centering
    \includegraphics[width=0.98\linewidth]{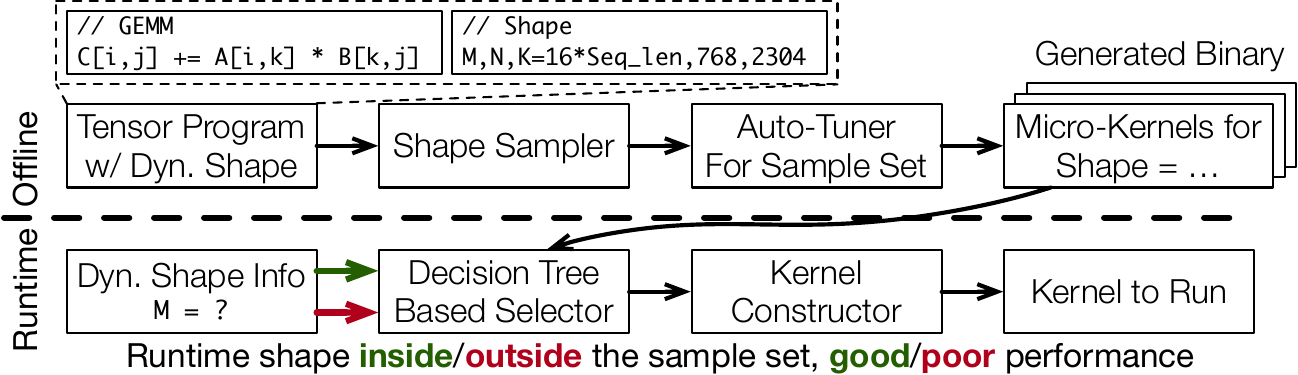}

        \caption{Existing sample-driven compilation workflow.}
        
    \label{fig:dietcode_workflow}

\end{figure}

\textbf{The intrinsic data format dynamism} is one of the fundamental driving forces demanding the integration of dynamic-shape tensor programs. For example, in natural language processing (NLP)~\cite{bert}, the inherent variability in sequence lengths is a characteristic feature that traditional static-shape tensor programs find challenging to accommodate. The diverse lengths of sentence tasks necessitate a flexible framework capable of dynamically adapting to varying input sizes. 
In computer vision (CV) tasks, conventional methods rely on fixed image sizes for tensor program input, which limits flexibility~\cite{mobilenet}. 
Recent developments~\cite{fast-rcnn, scale_fact_dect, dynamic_stride_net, dynamic_survey} have introduced dynamic-shape tensors as input, enhancing support for more advanced detection and tracking tasks.
Additionally, in the field of graph neural network (GNN)~\cite{gnn_survey, GNNAdvisor}, the dynamic nature of graph structures, marked by varying numbers of vertices and edges, necessitates the use of dynamic-shape tensor programs. These examples highlight the essential need for systems that seamlessly manage the intrinsic variability in diverse data formats.

\textbf{System execution and scheduling strategies} further underscore the importance of dynamic-shape tensor programs. For example, dynamic adjustment of batch sizes in the execution of neural networks introduces variability that demands adaptability in the underlying tensor program~\cite{dvabatch,lazybatch,orca}. The ability to efficiently handle varying batch sizes is crucial for optimizing resource utilization and achieving optimal performance in real-world applications.

\subsection{Limitations of Sample-Driven Approach}
\label{sec:mot:sample}

Many optimizations have been proposed for dynamic-shape tensor programs, all following a similar workflow~\cite{nimble,dietcode}. We term this methodology the \textbf{sample-driven} approach.
In this subsection, we first demonstrate its workflow and then discuss the inherent limitations.

As depicted in \autoref{fig:dietcode_workflow}, the current approach typically employs a sample list to depict the dynamic-shape parameters, combining the preknown static shape and the tensor program to staticize the dynamic-shape tensor program into a shape-generic search space.
An auto-tuning module, adapted from the static-shape compiler\cite{tvm,ansor,tensorir}, is subsequently utilized.
The auto-tuning module takes the tensor program with unspecified tile parameters, proceeding to investigate the high-performance tiling configurations within the comprehensive search space. This process produces fine-tuned micro-kernels for each input sample. At runtime, a decision-tree-based selector is employed to choose the appropriate micro-kernel based on the runtime shape. 
The kernel constructor finalizes the process by setting kernel launch parameters and incorporating padding to extend the applicability of these micro-kernels to runtime shapes not included in the sample list, utilizing the pre-compiled micro-kernels for runtime execution.

\begin{figure}[t]
    \centering

    \includegraphics[width=0.98\linewidth]{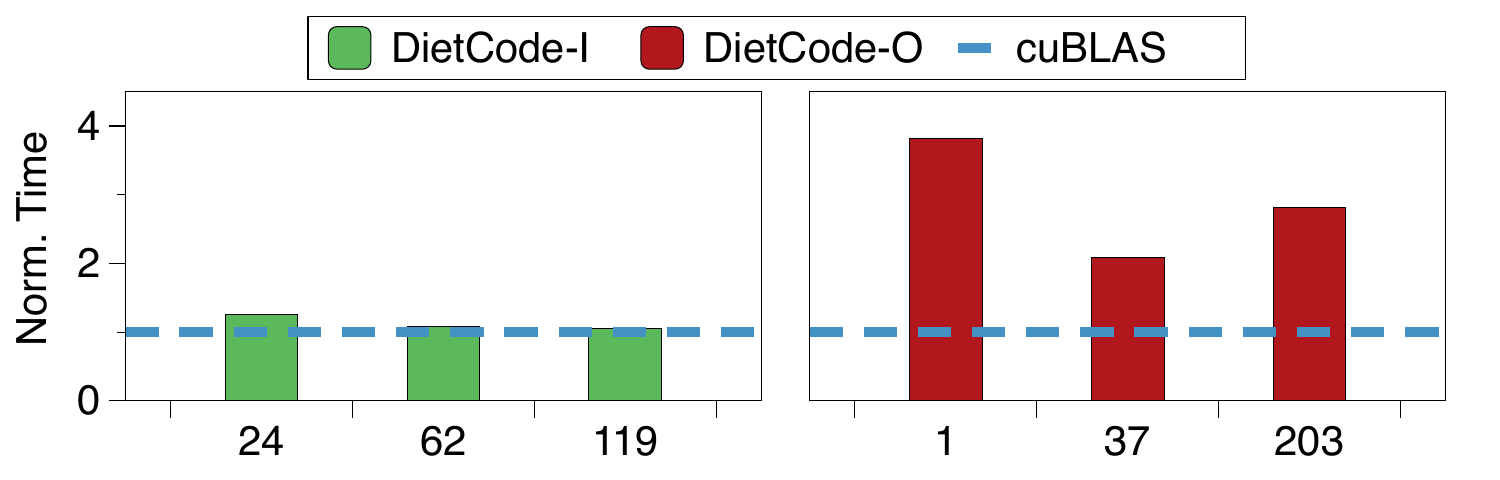}

    \caption{\small Comparing DietCode and cuBLAS over various sequence lengths on A100 GPU.
    `DietCode-I' and `DietCode-O' represent DietCode's dynamic input configurations inside and outside the tuning sample list, respectively.}
    \label{fig:mot:dietcode_rst}

\end{figure}

However, this approach faces significant limitations due to its reliance on a predetermined sample list for dynamic-shape parameters. This limitation restricts the compiler's flexibility and overall functionality. In situations where input shapes are not included in the sample list, this sample-driven dynamic compilation optimization does not consistently provide high-performance computing support. This is a critical shortfall, as the diversity and variability of real-world data often extend beyond the scope of the predefined sample list.

To empirically validate this limitation, we conduct an experiment targeting the first general matrix multiply (GEMM) operation of the Bert model~\cite{bert}. 
This GEMM operation entails the multiplication of two matrices, \(A\) and \(B\).
In this context, \(M\) denotes the number of rows in matrix \(A\), computed as the product of batch size and sequence length. \(N\) represents the number of columns in matrix \(B\), fixed at 768, and \(K\) corresponds to the number of columns in matrix \(A\) (and rows in matrix \(B\)), fixed at 2304.
Utilizing DietCode's default sample configuration, tests are performed with a fixed batch size of 16 and sequence length varying from 5 to 128 in the increment step of 19.

\autoref{fig:mot:dietcode_rst} shows that DietCode exhibits a significant performance discrepancy for dynamic parameters not included in the sample list, as compared to the results achieved with the vendor's library cuBLAS~\cite{cublas}.
This is due to the absence of specifically optimized micro-kernels for these shapes and the increased inefficiency from padding loss. 
Moreover, this method's rigidity becomes evident when considering the need to modify the sample list to accommodate various computational scenarios. Such modifications necessitate re-tuning of the system, which worsens its inflexibility.

\subsection{Hardware-Driven Approach: Opportunities and Challenges}\label{subsec:mot:opportunity}

\begin{figure}[t]
    \centering
    \includegraphics[width=0.95\linewidth]{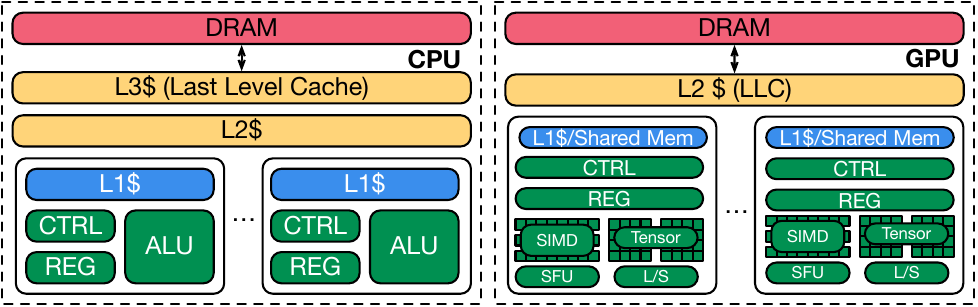}

    \caption{CPU/GPU Diagram.}
    
    \label{fig:cpu_gpu}

\end{figure}

The mismatch between runtime-used and offline-sampled shapes diminishes the efficiency of traditional sample-driven methodology in dynamic-shape compilation.
Such methodologies not only fall short in delivering high performance but also result in considerable tuning overhead.
However, we observe that generating the sample list for tuning micro-kernels is not mandatory for supporting the execution of a dynamic tensor program.
Sample-driven methodology treats the hardware platform as a black box, where micro-kernels are only tuned based on performance feedback of sampled shapes. This approach overlooks the rich vein of prior knowledge available within the hardware itself.

\autoref{fig:cpu_gpu} demonstrates the architecture of current mainstream hardware deployments, such as central processing units (CPUs)~\cite{intel_xeon_whitepaper} and graphics processing units (GPUs)~\cite{a100_whitepaper}.
There exist inherent similarities among these hardware, each of which has a hierarchical structure.
This hierarchy is distinctly multi-level, wherein each level comprises a predetermined quantity of computational or storage units. For instance, each CPU core has its own L1 cache and ALUs, while all the CPU cores share the L2/L3 cache and DRAM. As for GPUs, each streaming multiprocessor (SM) also has its own computing units (CUDA cores, Tensor Cores) and L1 Cache, while all SMs share the L2 cache and DRAM.
At each level, the resources available for executing a dynamic tensor program operator are constrained by the inherent limitations of the hardware design.

\begin{figure}[t]
    \centering

    \subfloat[CPU.]{
        \begin{minipage}[b]{0.99\linewidth}
        
            \centering
            \includegraphics[trim=0 0 0 0, clip, width=0.98\linewidth]{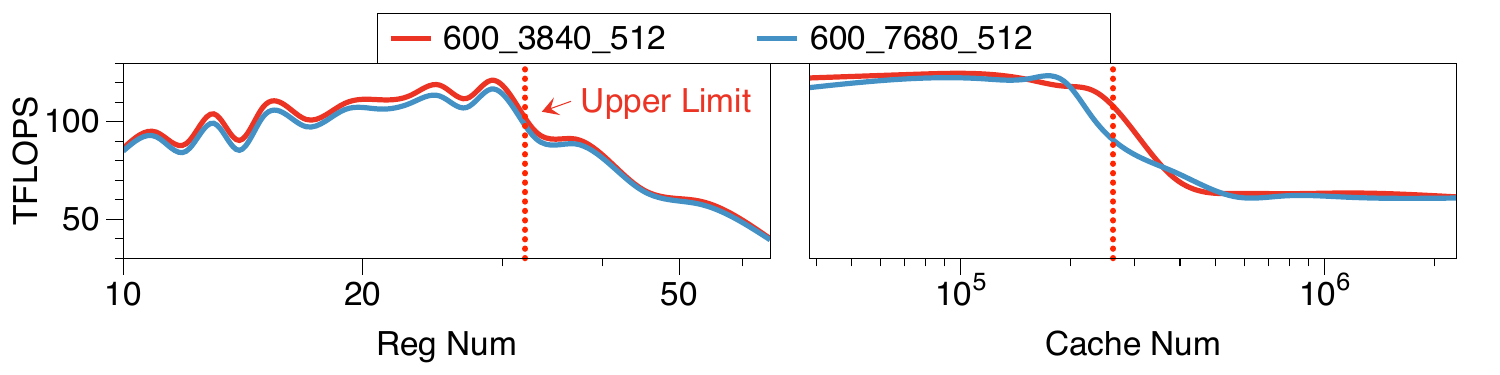}
            
        \end{minipage}
    }

    \subfloat[GPU.]{
        \begin{minipage}[b]{0.99\linewidth}
        
            \centering
            \includegraphics[trim=0 0 0 0, clip, width=0.98\linewidth]{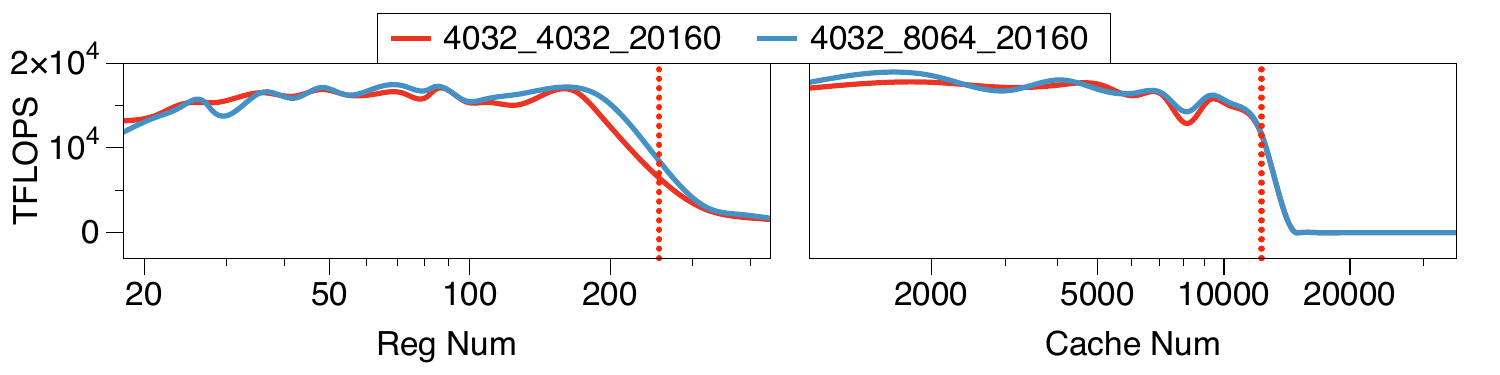}
            
        \end{minipage}
    }

    \caption{\small GEMM performance across different hardware resource usages on 8255c CPU and A100 GPU. Legend indicates corresponding GEMM parameters M, N, and K.}
    \label{fig:mot:hw_config}

\end{figure}

To investigate the impact of hardware limits on kernel performance, we experiment by generating configurations with different resource usages for the common-used matrix multiplication. 
We collect these configurations during Ansor~\cite{ansor}'s tuning process.
\autoref{fig:mot:hw_config} shows the floating point operations per second (FLOPS) and the corresponding resource usage of the collected configurations.
As observed, the operator performance seriously declines as the used resources exceed the hardware's upper limit.
This implies that configurations with hardware-unfriendly parameters consistently underperformed, allowing us to preemptively prune inefficient configurations from the large strategy space.
Consequently, this approach maintains a streamlined strategy space for micro-kernel generation and eliminates the need to create a predefined sample list.

Moreover, the dependency on hardware hierarchy motivates us to generate the micro-kernels in a bottom-up manner.
As the hardware hierarchy enables us to prune inefficient configurations based on the hardware limits level by level, only a limited number of micro-kernels is ultimately kept to support the dynamism of the tensor program.
This approach enables comprehensive support for all potential input shapes in dynamic tensor programming.

\paragraph{Summary} 
In the light of utilizing the inherent nature of hardware structure, we can not only optimize but also fundamentally rethink our approach to compilation strategy. This observation motivated us to design a hardware-aware dynamic-shape compiler that facilitates highly efficient computational performance while maintaining flexibility.

\section{Overview of \proj{}}
\label{sec:overview}

In this section, we describe the workflow of \proj{}, highlighting its key idea and optimization flow. \proj{} is an advanced operator-level compiler explicitly designed for optimizing the computation of dynamic-shape tensor programs. 

\paragraph{Key Idea.}
The key idea of \proj{} is its strategic use of hardware features to develop a hierarchical optimization flow, seamlessly integrating both offline and runtime stages.
Initially, \proj{} systematically decomposes dynamic-shape tensor programs into multi-level subtasks through a unified abstraction, \abst{}.
Each hierarchical level leverages hardware parameters to create micro-kernels tailored to specific hardware needs.
The process begins with the initial construction of micro-kernels at the lowest level, progressing to the final kernel selection during the runtime stage when shape information becomes available.
The hardware-aware approach of \proj{} facilitates a bottom-up, multi-level compilation workflow.
This approach leads to efficient optimization with lower runtime overhead and improved performance, which is ideal for dynamic-shape tensor programs.

\begin{figure}[t]
    \centering
    \includegraphics[width=0.98\linewidth]{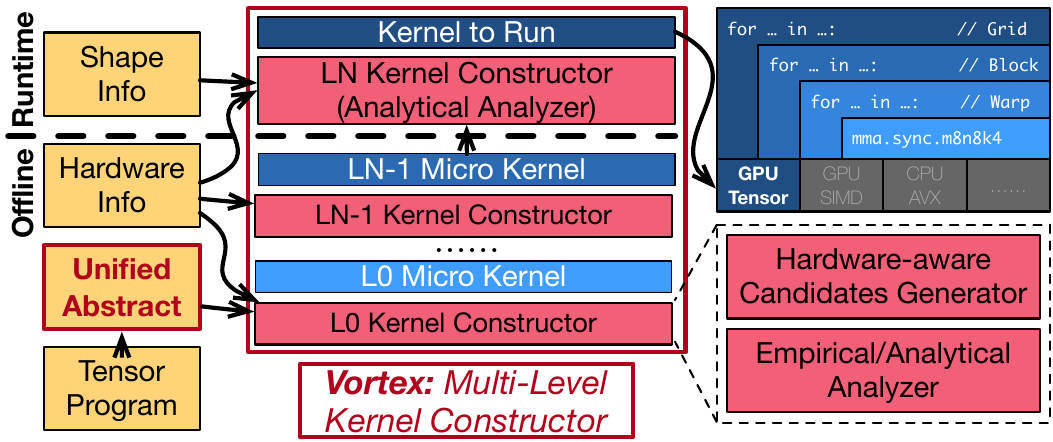}

    \caption{Design overview of \proj{}.} 
    
    \label{fig:overview}

\end{figure}

\paragraph{Optimization Flow.}
\autoref{fig:overview} details the methodological framework of \proj{}. 
In the offline stage, for a given tensor program, \proj{} employs a top-down abstraction to transform tensor programs into a structured, hierarchical format that mirrors the hardware’s structure, which aligns the strategy space with hardware architecture.
For instance, the upper-right section of \autoref{fig:overview} illustrates the tensor program abstraction of a GPU-based kernel. Here, we can seamlessly align the terms ``Grid'', ``Block'', and ``Warp'' with the device, SM, and Tensor Cores, respectively.
Then \proj{} employs a hardware-aware multi-level kernel constructor to develop a suite of hardware-friendly micro-kernels from the bottom level to the top.
At each level, \proj{} utilizes a micro-kernel generator to identify candidates, then employs empirical or analytical analyzers to fine-tune the micro-kernels' implementation.
This approach offers a well-balanced compromise between performance and overhead, tailored specifically to each hardware tier. 
Then \proj{} generates a set of micro-kernels that are highly compatible with the hardware in each level.
During the runtime stage, \proj{} employs a streamlined analytical module. This module quickly selects the appropriate candidate for the runtime input shapes, facilitating efficient execution on the hardware.

The benefits of this hardware-aware optimization workflow are two-fold. Primarily, \proj{} can generate hardware-friendly kernels that are universally capable of delivering high-performance computational support.
Furthermore, the offline compilation phase's independence from runtime shape samples significantly broadens the scope of support \proj{} can offer, as it reduces constraints and enhances flexibility.

\paragraph{Summary.}
In summary, \proj{} presents an effective and efficient solution to the challenges of dynamic-shape tensor program computation. Its hardware-aware and multi-level process, blending offline preparation with runtime efficiency, ensures the system's adaptability to complex dynamic-shape computational scenarios, guaranteeing high-performance support while maintaining low runtime overhead.

\begin{figure}[t]
    \centering
    \includegraphics[width=0.98\linewidth]{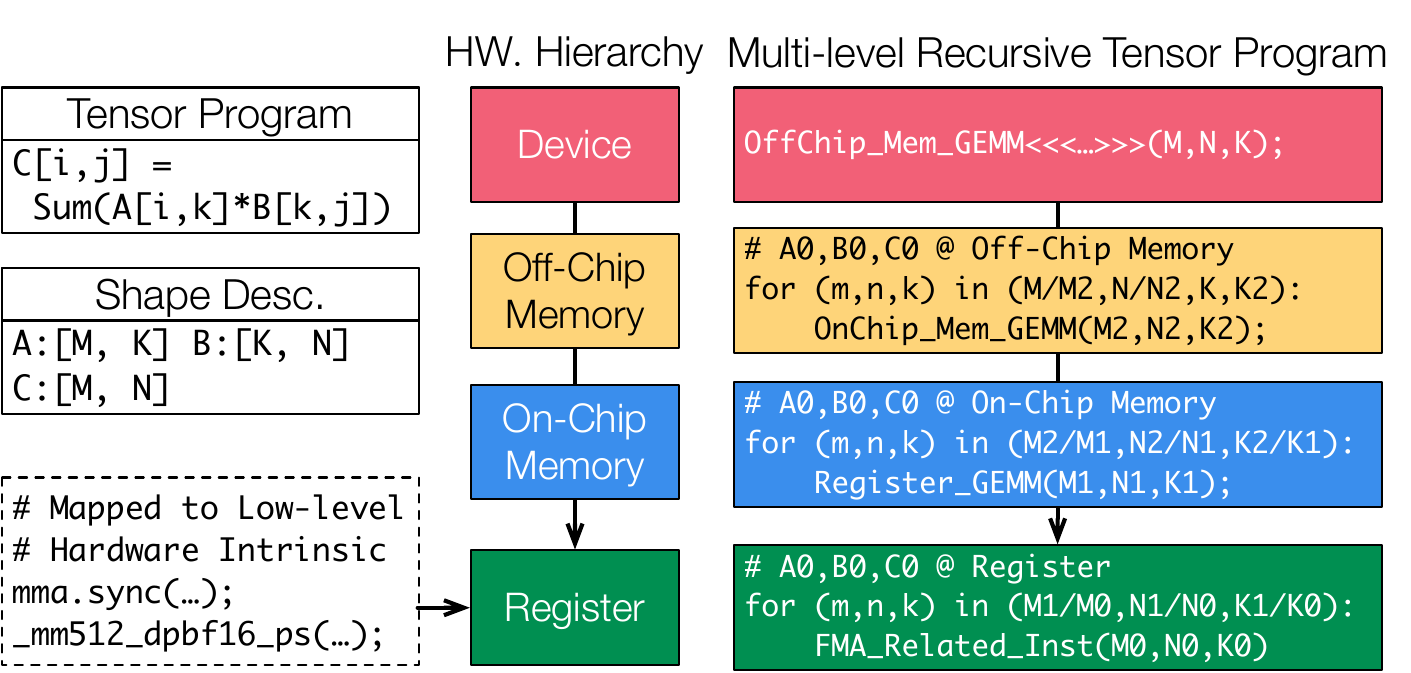}

    \caption{Recursive execution pattern of the \textit{GEMM} operator across hardware hierarchy levels.}
    \label{fig:multi_level_mapping}

\end{figure}

\section{Strategy Space Hierarchization in \proj{}}

In this section, we introduce the key of \proj{}'s workflow: strategy space hierarchization via top-down recursive decomposition. We first start with the \textit{GEMM} as an example to illustrate the recursive execution pattern inherent in tensor programs. 
This example lays the foundation for presenting our innovative unified abstraction, \abst{}, designed to accommodate the recursive nature prevalent in diverse tensor programs and across multiple hardware hierarchies.

\subsection{Top-Down Recursive Notation}

In this subsection, we use the example of \textit{GEMM} tensor program on GPUs as a paradigmatic example of recursive execution patterns on hierarchical hardware platforms. 
\textit{GEMM}, as a classic operator in deep learning, is mathematically defined as \( C = A \times B \), where \( A \) and \( B \) are the input matrices and \( C \) is the output matrix.

As shown in \autoref{fig:multi_level_mapping}, it is natural to use recursive for-loops to represent \textit{GEMM} execution flows. This approach breaks down the \textit{GEMM} tensor program into a series of recursive loops set. These recursive loop sets elegantly map the high-level \textit{GEMM} tensor program onto specific hardware levels, ranging from off-chip to on-chip memory and registers. 
Specifically, the uppermost recursion, \textit{OffChip\_Mem\_GEMM}, processes end-to-end matrix blocks in off-chip memory. At intermediate levels, \textit{OnChip\_Mem\_GEMM} handles smaller matrix blocks in on-chip memory. The innermost recursion, \textit{Register\_GEMM}, focuses on individual elements in registers, utilizing specific calculation instructions, such as \textit{Fused-Multiply-Add (FMA)}~\cite{a100_whitepaper} instructions on GPUs.

This top-down recursive approach is intuitive and crucial for optimizing tensor programs on modern hardware architectures.
It gives us an opportunity to decouple optimization space between different layers from each other, ensuring that each recursive layer is finely tuned to the unique capabilities and limitations of the corresponding hardware level, whether it be off-chip memory, on-chip memory, or registers.
Furthermore, this approach provides a clear structure for optimizing different aspects of the tensor program independently at each level.

\subsection{Unified Abstraction Design}

\begin{algorithm}[t]
\caption{Unified Recursive Abstraction}
\label{algo:unified_abst}
\small

\algtext*{EndWhile}
\algtext*{EndIf}
\algtext*{EndFor}

\begin{algorithmic}[1]
    \Procedure{rKernel}{$L, PL, TSL, TRL, LF, SF$}
        \State \textit{// L: Current hierarchical layer}
        \State \textit{// PL: Set of parallel loops}
        \State \textit{// TSL: Set of temporal spatial loops}
        \State \textit{// TRL: Set of temporal reduction loops}
        \For{each parallel loop $p$ in $PL[L]$}
            \For{each temporal spatial loop $ts$ in $TSL[L]$}
                \For{each temporal reduction loop $tr$ in $TRL[L]$}
                    \State \Call{Load\_Func}{$L, p, ts, tr$}
                    \State \Call{rKernel}{$L-1, PL, TSL, TRL, LF, SF$}
                \EndFor
                \State \Call{Store\_Func}{$L, p, ts$}
            \EndFor
        \EndFor
    \EndProcedure
\end{algorithmic}
\end{algorithm}

Recognizing the hierarchical nature of tensor program execution, we also acknowledge the variations between different hardware and tensor programs. 
Specifically, CPUs and GPUs exhibit distinct computation modes and memory access controls, where CPUs are optimized for multi-threaded parallelism, and GPUs excel in Warp, CTA, and Grid-level parallelism.
Moreover, different tensor programs demonstrate unique loop patterns; for instance, the loop characteristics of \textit{Convolution} markedly differ from those in \textit{GEMM}.

These variations inspire our design of a unified abstraction, which delineates a universal approach for representing tensor program executions across various hardware platforms.
\autoref{algo:unified_abst} elaborates on this abstraction.
It maintains the layer-wise recursive structure as demonstrated in \textit{GEMM} (\autoref{fig:multi_level_mapping}) as the core, and enables custom loop mapping and execution stages for various tensor programs.

To achieve a universal and customizable representation, we classify loops within each hierarchical level into three distinct sets in \autoref{algo:unified_abst}. 
The \textit{Parallel Loop Set} is designed for parallel execution; the \textit{Temporal Spatial Loop Set} manages temporal non-reduction operations; and the \textit{Temporal Reduction Loop Set} focuses on temporal reduction operations.
Each level, identified as level N, abstracts the execution into three stages: \textit{Load}, \textit{rKernel(N-1)}, and \textit{Store}. 
These stages serve as flexible interfaces, allowing for tailored execution based on the specific requirements of each hardware level.

\begin{table}[t]
\caption{\small Complete representation for different hardware, levels via \abst{} abstraction. `-' refers to `No Parallel Binding' in the `Parallel Binding' column and `No Operation' elsewhere.}
\label{tab:abst_hw_mapping}

\small
\centering
\resizebox{0.49\textwidth}{!}{

\begin{tabular}{|c|c|l|l|l|l|}
\hline
\textbf{HW.} & \textbf{Level} & \begin{tabular}[c]{@{}l@{}}\textbf{Parallel} \\ \textbf{Binding}\end{tabular} & \textbf{Load } & \begin{tabular}[c]{@{}l@{}}\textbf{Lower Level} \\ \textbf{rKernel}\end{tabular} & \textbf{Store} \\ 
\hline
\multirow{3}{*}[-1em]{CPU}                                                   
    & 0    & -                 & \begin{tabular}[c]{@{}l@{}}CacheBuf \(\rightarrow\) Reg \\ \textbf{or}   GlobalMem\(\rightarrow\) Reg\end{tabular}   & ALU Calc.        & \begin{tabular}[c]{@{}l@{}}Reg \(\rightarrow\)  CacheBuf\\ \textbf{or}  Reg \(\rightarrow\)  GlobalMem\end{tabular}   \\ \cline{2-6}

    & 1    & Thread                 & \begin{tabular}[c]{@{}l@{}}GlobalMem \(\rightarrow\) CacheBuf \\ \textbf{or}  - \end{tabular} & L1 rKernel       & \begin{tabular}[c]{@{}l@{}}CacheBuf \(\rightarrow\) GlobalMem \\ \textbf{or}  - \end{tabular} \\ \cline{2-6} 
    
    & 2    & Process       & -                 & L2 rKernel       & -                               \\ 
    
\hline
\multirow{3}{*}[-1em]{GPU} 
    
    & 0    & Warp              & \begin{tabular}[c]{@{}l@{}} SharedMem\(\rightarrow\) Reg \\  \textbf{or} GlobalMem\(\rightarrow\) Reg\end{tabular}     & \begin{tabular}[c]{@{}l@{}} Cuda/Tensor \\  Core Calc. \end{tabular} & \begin{tabular}[c]{@{}l@{}}Reg \(\rightarrow\)SharedMem  \\ \textbf{or}  Reg \(\rightarrow\) GlobalMem\end{tabular}     \\ \cline{2-6} 
    
    & 1   & CTA               & \begin{tabular}[c]{@{}l@{}}GlobalMem \(\rightarrow\) SharedMem \\  \textbf{or} - \end{tabular}  & L0 rKernel       & \begin{tabular}[c]{@{}l@{}}SharedMem \(\rightarrow\) Global \\ \textbf{or}  - \end{tabular}  \\ \cline{2-6} 
    
    & 2   & Grid              & -                               & L1 rKernel       & -                               \\ 
\hline
\end{tabular}
}

\end{table}

\autoref{tab:abst_hw_mapping} illustrates how this unified abstraction rKernel is implemented across different hardware configurations, highlighting its versatility. 
Our focus lies in scrutinizing recursive execution patterns among different hierarchies.
For CPUs, 
at the lowest level (L0), the \abst{} abstraction allows for direct data transfer from Global memory to Registers or from CacheBuffer to Registers, depending on the specific needs of the computation. 
A ``CacheBuffer" is defined as a memory buffer, sized within the L2 cache limits, to ensure consistent caching of its contents in the L2 cache for efficient data access and processing~\cite{onednn}. 
Additionally, store operations at this level also reflect this adaptability, offering the choice of transferring data back to Global memory or CacheBuffer. As we progress to level L1, the abstraction provides options for either transferring data from Global memory to CacheBuffer or performing no operation, signifying a versatile approach to data handling. The highest level (L2) in CPUs focuses on the multi-thread mechanism at the process level, capitalizing on the CPU's capabilities for multi-core parallel processing.

Similarly, for GPUs, including both Cuda Cores and Tensor Cores, \abst{} adapts to different operational requirements. At L0, there's an option for loading data either from Global memory or Shared Memory to Registers, and similarly, storing data either back to Global memory or to Shared Memory. This flexibility is crucial for optimizing memory utilization in the highly parallel environments of GPUs. At the L1 level, similar to CPUs, the abstraction can facilitate data transfer from Global to Shared Memory or no operation, enabling efficient resource management. The L2 level focuses on Grid-level operations, enhancing the scalability across the GPU's multiple streaming multiprocessor (SM) architecture.

\abst{} achieve a hierarchy abstraction of execution patterns that apply universally to various tensor programs and hardware types. This approach ensures a tailored strategy space for each hardware hierarchy level and facilitates the universal optimizations for dynamic-shape tensor programs.

\section{Detailed Designs at Each Level }

In this section, we thoroughly explore \proj{}'s detailed designs at each hierarchical level. Initially, \proj{} utilizes an effective approach to generate hardware-aware candidates at each level.
Following this, a hybrid analytical-empirical analyzer is deployed to discern the high-performance implementation for each candidate.

\subsection{Bottom-up Hardware-aware Candidates Generator}
\label{subsection:candidate_generate}

This subsection introduces an innovative bottom-up method for generating candidates tailored to align with top-down recursive abstraction. 
This method becomes particularly crucial when execution parameter information is lacking, posing a significant challenge in every hierarchical layer for identifying suitable shape candidates for micro-kernel creation. 
Our approach centers on two key processes:

\begin{figure}[t]
    \centering
    \includegraphics[width=0.9\linewidth]{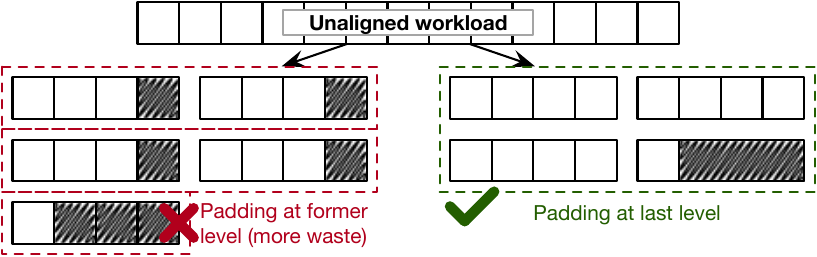}

    \caption{\small Comparison of different padding patterns and their corresponding waste scenarios.}
    \label{fig:padding}

\end{figure}

Firstly, we utilize the hardware's parameter information to determine constraints for the candidate range. As highlighted in \S\ref{subsec:mot:opportunity}, our empirical studies have identified a notable decrease in hardware execution efficiency when the utilization at any hardware level is extremely low or high. This insight allows us to deduce a feasible range for candidate shapes, based on hardware utilization metrics.

Secondly, we follow a key design principle: ensuring the shape size of candidates in an upper layer is an integer multiple of the shape size in the lower layer. This approach aims to minimize padding loss during the construction of micro-kernels. As shown in \autoref{fig:padding}, if the sizes at one level are not multiples of those below, it results in more padding losses and inefficiencies at higher levels. Conversely, constructing candidates as integer multiples from one level to the next predominantly confines padding loss only to the outermost execution level, adhering to runtime requirements.

\begin{algorithm}[t]
\caption{Candidates Generation Algorithm.}
\label{algo:sieve_based_param_determination_specific_layer}
\small

\algtext*{EndWhile}
\algtext*{EndIf}
\algtext*{EndFor}

\begin{algorithmic}[1]
    
    \Function{GenerateCandidatesForLayer}{$L$}
        
        \State $hwInfo \gets$ \Call{GetHardwareInfo}{$L$}

        \State \textit{// Determine by hardware resource limitation}
        \State $cands \gets$ \Call{InitCands}{$hwInfo$}
        \If{$L = 0$}
            
            \State $cands \gets$ \Call{FilterByISA}{$cands$}
        \Else
            \State $prevCands \gets$ \Call{GetPrevLayerCands}{$L - 1$}
            
            \State $cands \gets$ \Call{FilterByMultiples}{$cands$, $prevCands$}
        \EndIf
        \State \Return $cands$
    \EndFunction

    \Function{FilterByISA}{$cands$}
        \State $filtered \gets \emptyset$
        
        \For{$cand \in cands$}
            \If{\Call{IsCompatible}{cand}}
                
                \State $filtered$.add($cand$)
            \EndIf
        \EndFor
        \State \Return $filtered$
    \EndFunction
    \Function{FilterByMultiples}{$cands, prevCands$}
        \State $filtered \gets \emptyset$
        \State $map \gets$ an empty map
        \For{$prev \in prevCands$}
            \State $multiples \gets$ \Call{GenerateMultiples}{$prev$, $cands$}
            \For{$multiple \in multiples$}
                
                \State $filtered$.add($multiple$)

                \State $map[multiple]$.append($prev$)
                
            \EndFor
        \EndFor
        \State \Return $filtered, map$
    \EndFunction

\end{algorithmic}
\end{algorithm}

The overall process is detailed in \autoref{algo:sieve_based_param_determination_specific_layer}. The core function, \textit{GenerateCandidatesForLayer}, is designed to operate distinctly based on the hierarchical layer it addresses. 
It begins by acquiring the hardware specifications for a given layer (denoted as L). This is achieved through the \textit{GetHardwareInfo} and \textit{InitCands} function, which retrieves essential hardware constraints that critically influence the parameter space. 
To validate candidate feasibility, the key is assessing memory usage against layer-specific limits and accounting for hardware constraints, such as a GPU's 1024 threads-per-block maximum.
For the initial layer (L = 0), the function employs \textit{FilterByISA} to refine the candidate set according to the hardware platform's Instruction Set Architecture (ISA) compatibility. For instance, on Intel CPUs, the \textit{FilterByISA} function considers the granularity constraints of AVX512~\cite{intel_xeon_whitepaper}. Similarly, on GPU with Tensor Cores, the function assesses the constraints imposed by the Matrix Multiply-Accumulate (MMA) instruction~\cite{a100_whitepaper}. These considerations ensure alignment with hardware capabilities.

Our algorithm utilizes \textit{FilterByMultiples}, a method inspired by the classic sieve approach~\cite{sieve}, to filter candidates.
This function iteratively processes previous layer candidates (prevCands), generating multiples constrained via \textit{GenerateMultiples} within the current layer's candidate range. 
This approach ensures comprehensive exploration of viable parameter sets and maintains filtering efficiency. 
Additionally, we employ a mapping mechanism that uses a table to record the links between each candidate in the current layer and its possible match candidates in the previous layer, which is crucial for the subsequent analysis module.

\subsection{Hybrid Analytical-Empirical Analyzer}

In this subsection, we present a novel analyzing method which effectively incorporates analytical and empirical methodologies. 
This approach is designed to optimize the trade-off between efficiency and accuracy in strategy analysis.

\paragraph{Goal of the Analyzer.}
Our analyzer aims to identify suitable candidates at each layer.
Using \autoref{algo:sieve_based_param_determination_specific_layer}, we construct a map that describes the connections between candidates in connected layers. However, a single candidate can map to multiple lower-level candidates in this table. Each mapping corresponds to a unique implementation of the scheduling strategy, necessitating a thorough evaluation of the performance variability among these strategies. This task of identifying the optimal strategy $c^*$ from a set $S$ at a given hardware level $L$ can be defined as an optimization problem:

\begin{equation}
c^*=\underset{s \in S}{\arg \min } \, \operatorname{Cost}(s, L)
\end{equation}

The analyzer's primary objective is to identify a cost-effective and efficient strategy that aligns with the requirements at both offline and runtime stages. Concurrently, it is crucial to be aware of the time overhead associated with cost analysis, since excessive overhead, particularly during runtime, is clearly unacceptable. This necessitates a sophisticated design approach, blending empirical profiling on actual hardware with comprehensive theoretical analysis.

\begin{figure}[t]
    \centering
    \includegraphics[width=0.8\linewidth]{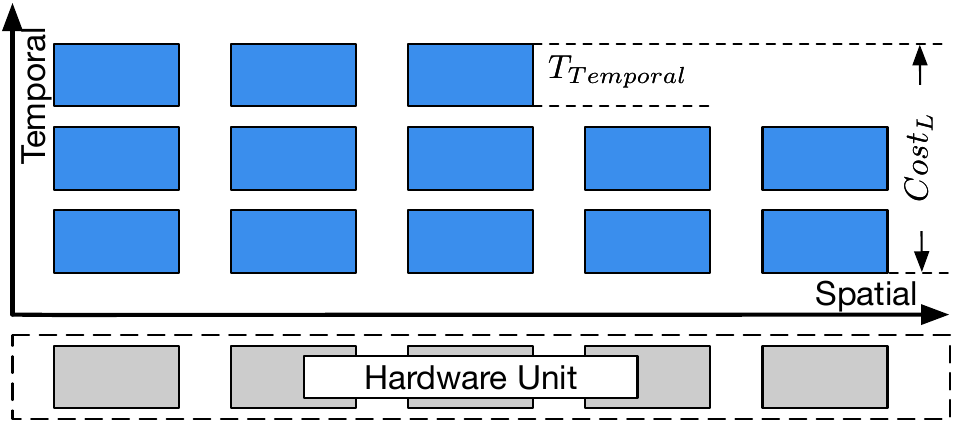}

    \caption{An Illustration of execution abstraction and associated analytical model.}
    \label{fig:cost_model}

\end{figure}

\paragraph{Analytical Cost Model.}
We build an analytical cost model, a theoretical framework to predict the costs of different candidate implementations. The analytical cost model encapsulates the execution time based on algorithmic complexity and hardware specifications. As shown in \autoref{fig:cost_model}, we central to this model in two determinants: \( Spatial \), representing the amplification factor due to parallel loops executed across different hardware units, and \( Temporal \), relating to the execution process of serially dependent loop operations.

The temporal execution cost, \( T_{temporal} \), is carefully crafted to encapsulate the intricacies of pipeline execution within serial loops. It is quantified as follows:

\begin{equation}
\begin{split}
    T_{temporal} = &  T_{Load} + \bigl(\text{sizeof}(\text{TemporalLoop}) - 1\bigr) \\ 
    & \times \max\bigl(T_{Load}, Cost_{L-1}\bigr) + Cost_{L-1} + T_{Store}
\end{split}
\end{equation}

In this equation, \( T_{Load} \) and \( T_{Store} \) are the costs taken to load and store data, respectively. They are calculated based on the amount of data moved at the current layer divided by the memory bandwidth at that layer. 
\( Cost_{L-1} \) represents the cost taken by the micro-kernel computation at the lower level.
This equation accounts for the loop's iteration count and juxtaposes the data load time against the execution span of a reduced kernel operation, thereby emulating the pipeline's potential latency bottlenecks.

In parallel processing, the cost is modulated by the parallel loop's scale relative to the hardware's unit capacity:
\begin{equation}
    F_{parallel} = \left\lceil \frac{\text{sizeof}(\text{ParallelLoop})}{ | \text{HardwareUnit} | } \right\rceil
\end{equation}

This quantifies the hardware's parallel processing ability, scaling execution cost with the parallel loop size. The overall strategy cost at layer L, \(Cost_{L} \), is the product of the parallel execution cost and the temporal execution cost, capturing the total time required for the execution of a computational task across various layers:

\begin{equation}
    Cost_{L} = F_{parallel} \times T_{temporal}
\end{equation}

Our analytical cost model faces a recursive complexity, needing $Cost_{L-1}$ for each level L. 
Furthermore, hardware optimizations such as instruction pipelining and out-of-order execution can lead to substantial inaccuracies in the cost model, posing a challenge in ensuring the precision of the analysis module~\cite{micro23_path}.

\paragraph{Hybrid Analyzer Design.}
Two key observations guide our design. Firstly, the bottom-up multi-level approach to kernel construction incrementally increases the number of candidates at higher layers. Secondly, unpredictable hardware-related scheduling, such as out-of-order execution, predominantly focuses on lower layers. These insights led to the development of our hybrid analytical-empirical analyzer.

The analyzer conducts empirical profiling on CPUs at level L0 and on GPUs at both L0 and L1 levels. For higher levels, it utilizes an analytical cost model. This hybrid system synergizes the efficiency of the analytical approach with the accuracy of empirical data, with the latter offering real-time performance insights to augment the analytical predictions. Importantly, all runtime analyses are conducted using the analytical model, ensuring a streamlined and low-overhead performance evaluation. 
The effectiveness of this hybrid methodology, in terms of performance and runtime overhead, is further
investigated in \S\ref{subsec:addition_analysis}. 
Overall, this hybrid approach is especially valuable in complex scenarios where theoretical models may be insufficient, ensuring both effectiveness and precision in the strategy analysis process.

\section{Implementation}
\label{sec:implementation}

In this section, we detail the implementation of \proj{}. We focus on demonstrating the code generation method, and the scheduling process during the runtime phase.

\begin{figure}[t]
    \centering

    \lstset{basicstyle=\footnotesize\ttfamily}
    
    \begin{lstlisting}[language=c]
enum ANALYZE_TYPE {empirical, analytical};
enum LOOP_TYPE {PL, TSL, TRL};
class axis;
class layer_meta_info {
  int    layer_depth;
  map<axis, LOOP_TYPE> loop_type;
  ANALYZE_TYPE  analyzer;
  
  func* load_func;
  func* store_func;
  func* compute_func;
};
    \end{lstlisting}

    \caption{The definition of \abst{}.}
    \label{fig:structure_abst}
    
\end{figure}

\subsection{Code Generation}

Despite the contrasting architectures of GPUs and CPUs, \proj{}'s abstraction consistently represents both, as exemplified in \autoref{tab:abst_hw_mapping}. One of the key aspects contributing to this universal representation is the definition of data structures.

The \abst{} data structure, as illustrated in \autoref{fig:structure_abst}, is a cornerstone of \proj{}, tailored to encapsulate and streamline the complex processes involved. At its core lies the \texttt{layer\_meta\_info} class, pivotal for orchestrating the optimization strategy of each hierarchical layer. Within this class, the \texttt{layer\_depth} attribute determines the layer's position within the hierarchical structure, which is a crucial factor for the recursive optimization process. 
Moreover, the \texttt{map<axis, LOOP\_TYPE>} provides a strategic mapping of loops to their respective types, including Spatial (S), Temporal Parallel (TP), and Temporal Reduction (TR). This nuanced approach to loop optimization aligns seamlessly with the unique characteristics of each tensor program.
The \texttt{ANALYZE\_TYPE enum}, encompassing empirical and analytical options, facilitates the selection of the appropriate optimization analysis method. 
Finally, the functional pointers—\texttt{load\_func}, \texttt{store\_func}, and \texttt{compute\_func}—are critical in dynamically managing the various stages of computation. This data structure encapsulates the necessary elements to navigate dynamic-shape tensor program optimization challenges adeptly.

\begin{figure}[t]
    \centering

    \lstset{basicstyle=\footnotesize\ttfamily}

    \begin{lstlisting}[language=c]
// Input IR, we omit block and grid for brevity
layer_meta_info gemm_tc_warp;
gemm_tc_warp.set(
  layer_depth  = 0,
  loop_type  = {"k0":TRL,"m0":TSL,"n0":TSL},
  cost_model = profiling;
  load_func  = load(shared_to_reg),
  store_func = store(reg_to_shared),
  compute_func = asm("mma.sync.m16n8k16")
);

// Output Generated Kernel 
dim3 grid(M/m_tile_grid, N/n_tile_grid);
gemm_tensor_core_grid<<<grid, thread>>>(...);

__global__ void gemm_tensor_core_grid(...) {
  __shared__ half A_buf[], B_buf[], C_buf[];
  for (k2 = 0; k2 < K; k2+=k_tile_grid) 
    for (k1 = 0; k1 < k2; k1+=k_tile_block) 
      asm("ld.global");  // Load A/B to A_buf/B_buf
      for (m0 = 0; m0 < m1; m0+=m_tile_warp) 
        for (n0 = 0; n0 < n1; n0+=n_tile_warp) 
          asm("ld.shared"); // Load A_buf/B_buf to A_reg/B_reg
          C_frag = 0;
          for (k0 = 0; k0 < k1; k0+=k_tile_warp) 
            asm("mma.sync.m16n8k16");
          C_frag += ...
          asm("st.shared"); // Store C_reg to C_buf
      asm("st.global");  // Store C_buf to C
}
    \end{lstlisting}

    \caption{An illustration of GPU GEMM code generation.}
    \label{fig:code_gen}

\end{figure}

\abst{} serves as a recursive-based template for dynamic tensor program. When dealing with a fixed hardware and a particular operator, it necessitates users to initialize different levels of \texttt{layer\_meta\_info}. 
The development effort required for this task is minimal. This is because, for both CPU and GPU, we set the hierarchy level to three, which does not impose a significant burden.
Furthermore, the dynamic parameters of the template enable consistent computational support for various runtime shapes. 
Owing to its versatility, development costs are apportioned across various computing scenarios.
We built \proj{} on top of TVM~\cite{tvm}. We harness the `tensorize' primitive to implement the load, store, and compute functions, and leverage TVM's robust and versatile code generation capabilities, we seamlessly target both CPU and GPU platforms.
In \autoref{fig:code_gen}, we illustrate a representative GEMM kernel code generation for GPUs, showcasing our methodology's depth and adaptability in optimizing dynamic-shape tensor programs for GPU architectures.

\subsection{Integration of Offline and Runtime }

The integration of offline and runtime components in \proj{} involves several key steps. During runtime, \proj{} employ analytical cost models to estimate the execution costs associated with different candidate solutions with runtime shape information. Subsequently, \proj{} select the most suitable micro-kernel candidates based on these optimal execution cost estimations.

Notably, our selection process accommodates the dynamic nature of hardware platforms.
For instance, in the case of GPUs, the presence of a larger MMA instruction~\cite{a100_whitepaper} padding in the tensor core necessitates adaptive hardware solutions.
We provide implementations for both CUDA cores and Tensor cores, allowing us to choose the appropriate backend hardware based on the runtime input shapes adaptively, further optimizing execution efficiency. The performance benefits of this adaptive strategy are thoroughly explored in \S\ref{subsec:addition_analysis}.
Additionally, by considering the computational shape of the selected micro-kernels in conjunction with the runtime shape, we collectively compute runtime-specific computational details, such as grid configurations. This comprehensive approach ensures the versatility and generality of our computational framework, facilitating efficient and adaptable runtime operations.

\begin{figure*}[t]
    \centering
    \subfloat[GEMM on CPU.]{

            \includegraphics[trim=0 0 0 0, clip, width=0.5\linewidth]{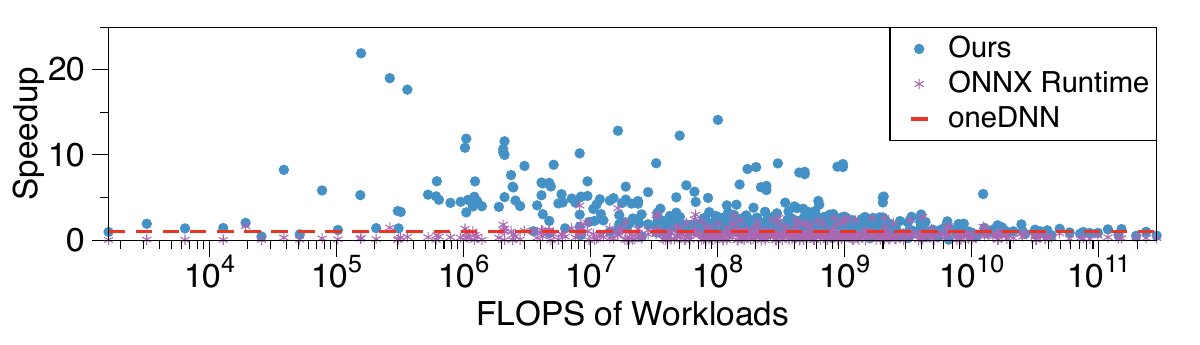}

    }
    \subfloat[Conv. on CPU.]{

            \includegraphics[trim=0 0 0 0, clip, width=0.5\linewidth]{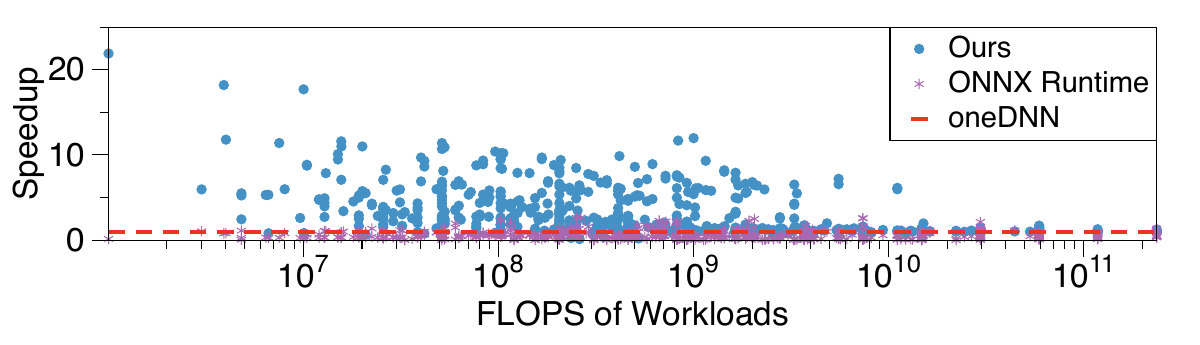}

    }

    \subfloat[GEMM on GPU with Tensor Core Enabled.]{
        \begin{minipage}[b]{0.50\linewidth}
        
            \centering
            
            \includegraphics[trim=0 0 0 0, clip, width=0.98\linewidth]{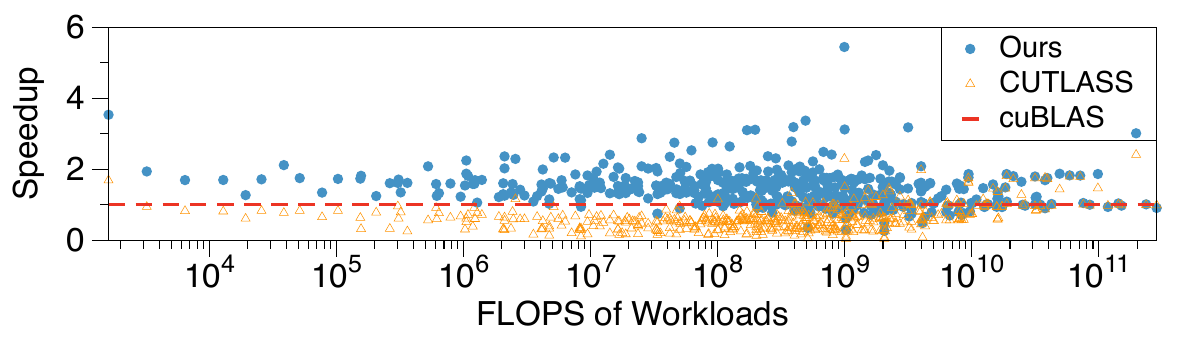}

        \end{minipage}
    }
    \subfloat[Conv. on GPU with Tensor Core Enabled.]{
        \begin{minipage}[b]{0.50\linewidth}
        
            \centering
            
            \includegraphics[trim=0 0 0 0, clip, width=0.98\linewidth]{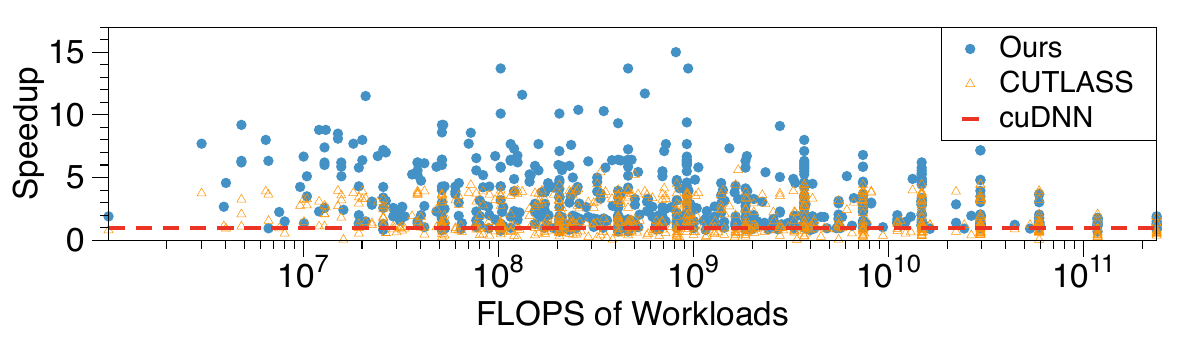}

        \end{minipage}
    }

    \subfloat[GEMM on GPU with Cuda Core Only.\label{subfig:op_benchmark_rst_gpu_cudacore}]
        {
        \begin{minipage}[b]{0.50\linewidth}
        
            \centering
            
            \includegraphics[trim=0 0 0 0, clip, width=0.98\linewidth]{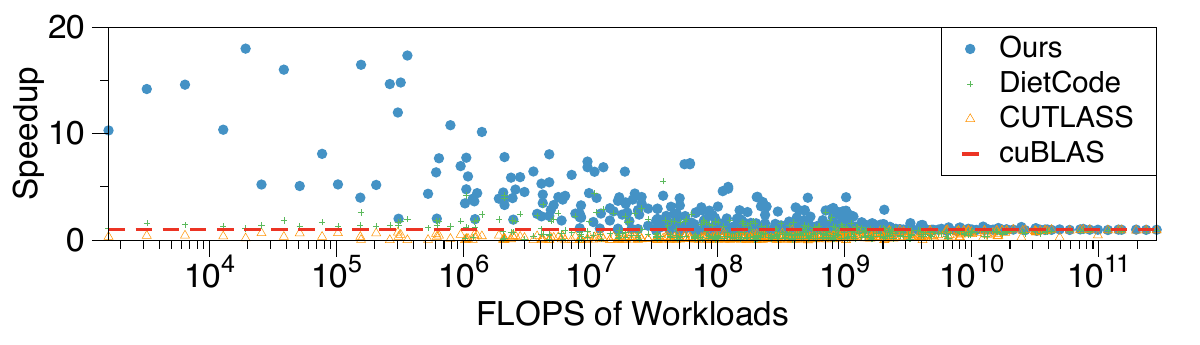}

        \end{minipage}
    }
    \subfloat[Conv. on GPU with Cuda Core Only.]{
        \begin{minipage}[b]{0.50\linewidth}
        
            \centering
            
            \includegraphics[trim=0 0 0 0, clip, width=0.98\linewidth]{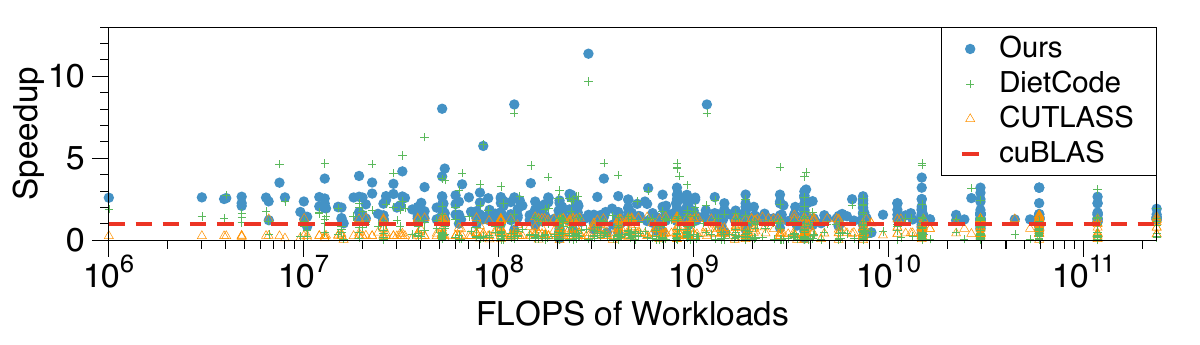}

        \end{minipage}
    }

    \caption{Performance results at operator-level. CPU results are normalized to oneDNN; GPU results are normalized to cuBLAS for GEMM and cuDNN for convolution, respectively. }
    \label{fig:op_benchmark_rst}
    
\end{figure*}

\section{Evaluation}
\label{sec:evaluation}

\subsection{Experimental Setup}

\paragraph{Platforms.} We evaluate \proj{} on two representative platforms: Intel 8255c CPU~\cite{intel_xeon_whitepaper} and Nvidia Ampere A100 GPU~\cite{a100_whitepaper}. 
For GPUs, we conduct evaluations using two computational modes: \textit{Tensor Core Enabled} mode with half-precision floating-point (FP16) data type, and \textit{Cuda Core Only} mode with single-precision floating-point (FP32) data type.
\autoref{tab:hardware} details our experimental platforms.

\begin{table}[t]
\centering

\caption{Hardware specifications.}
\label{tab:hardware}

\resizebox{0.49\textwidth}{!}{

\begin{tabular}{|c|c|c|}
\hline
\textbf{Hardware} & \textbf{Nvidia GPU}                                                                                                       & \textbf{Intel CPU}                                                                                                                           \\ \hline
Version           & Ampere A100 (108 SMs)                                                                                                     & Xeon 8255c (48 Cores)                                                                                                                        \\ \hline
Storage           & \begin{tabular}[c]{@{}c@{}}Global: 40G; L2 Cache:   40M;\\      Shared Memory: 48 K/SM; \\      Reg: 256K/SM\end{tabular} & \begin{tabular}[c]{@{}c@{}}Global: 250.53G; L3 Cache:   35.75M;\\      L2 Cache: 1M/Core; L1 Cache: 32K/Core;\\      Reg: 2K/Core\end{tabular} \\ \hline
Peak Flops        & \begin{tabular}[c]{@{}c@{}}CUDA Core: 19.5 TFlops;\\      Tensor Core: 312 TFlops\end{tabular}                               & 7344 GFlops                                                                                                                                  \\ \hline
OS                & CentOS Linux 8.4.2105                                                                                                     & CentOS Linux 8.6.2205                                                                                                                        \\ \hline
Software          & \begin{tabular}[c]{@{}c@{}}Driver Version: 450.156.00\\      CUDA Version: 11.8\\      cuDNN: 8.9.7.29\end{tabular}       & \begin{tabular}[c]{@{}c@{}}GCC: 10.2.0\\      LLVM: 15.0.3\end{tabular}                                                                      \\ \hline
\end{tabular}
}

\end{table}

\paragraph{Benchmarks.}
Our evaluation encompasses two benchmark categories: operator-level and model-level.
At the operator level, we gather 1197 different operator configurations from DeepBench~\cite{deepbench} and real-world models, covering various tasks like Transformer, CNN, and GNN (see \autoref{tbl:benchmark_gemm} and \autoref{tbl:benchmark_conv} for details). These configurations demonstrate variability in all dimensions and possess a wide dynamic range, making them highly representative. 
At the model level, we assess the performance of three Transformer-based language models (Bert~\cite{bert}, Bert-large~\cite{bert}, GPT2~\cite{gpt2}) and three computer vision models (AlexNet~\cite{alexnet}, ResNet~\cite{resnet}, GoogleNet~\cite{googleNet}) for end-to-end dynamic-shape neural network evaluation. To mirror real-world scenarios, we generate 17 sequence lengths ranging from 1 to 476 for language models. For CNN models, we configure batch sizes beginning with 1, and then incrementally step from 4 to 64 in multiples of 4.

\begin{table}[t]
    \centering

    \caption{Benchmarked GEMM with dynamic shapes.}
    \label{tbl:benchmark_gemm}
    
    \resizebox{0.49\textwidth}{!}{
    \begin{tabular}{|c|c|c|c|c|}
    \hline
    \textbf{Category}    & \textbf{M}                  & \textbf{N}              & \textbf{K}                & \textbf{\#Cases} \\ \hline
    DeepBench~\cite{deepbench} & {[}35, 8448{]}      & {[}1, 6000{]}   & {[}128, 500000{]} & 84           \\ \hline
    Transformer~\cite{bert,gpt2,huggingface}       & {[}1, 476{]}       & {[}768, 4096{]} & {[}768, 4096{]}   & 192          \\ \hline
    CNN~\cite{alexnet,vgg,googleNet,resnet}       & {[}1, 128{]}        & {[}80, 25088{]} & {[}10, 4096{]}    & 80          \\ \hline
    GNN~\cite{GCN,GAT,GNNAdvisor,pyg}       & {[}2708, 1888584{]} & {[}2, 121{]}    & {[}8, 3703{]}     & 150          \\ \hline
    \end{tabular}
    }
    
    \end{table}
    
    \begin{table}[t]
    \centering

    \caption{Benchmarked Convolution with dynamic shapes.}
    \label{tbl:benchmark_conv}

    \resizebox{0.49\textwidth}{!}{
    \begin{tabular}{|c|c|c|c|c|c|c|}
    \hline
    \textbf{Category} & \textbf{BS} & \textbf{Fmap} & \textbf{Filter} & \textbf{Cin} & \textbf{Cout} & \textbf{\#Cases} \\ \hline
    DeepBench~\cite{deepbench}       & {[}1,16{]}         & {[}7,700{]}        & {[}1,20{]}           & {[}1,2048{]} & {[}16,2048{]} & 107                   \\ \hline
    CNN~\cite{alexnet,resnet,vgg,googleNet}             & {[}1,64{]}         & {[}4,768{]}        & {[}1,11{]}           & {[}3,832{]} & {[}16,512{]}  & 584                   \\ \hline
    \end{tabular}
    }

    \end{table}

\paragraph{Baselines.}
We select various SOTA baselines, divided into two principal categories. The first category encompasses vendor-provided libraries, which are specialized libraries frequently utilized in neural network frameworks~\cite{pytorch}. 
For NVIDIA GPUs, our evaluation utilizes cuBLAS~\cite{cublas} for GEMM and cuDNN~\cite{cudnn} for convolution. We also evaluate CUTLASS~\cite{cutlass} for both tasks.
For Intel CPUs, we compare GEMM and convolution performance with oneDNN~\cite{onednn} and ONNX Runtime~\cite{onnxruntime}. 
The second category is dynamic-shape compilers, for which we select DietCode~\cite{dietcode}, the existing leading dynamic-shape tensor program compiler.

\begin{table}[b]

    \caption{Summary of operator-level speedups for \proj{} compared to various baselines across different setups.}
    \label{tbl:op_eval_rst}
    
    \resizebox{0.49\textwidth}{!}{
    \begin{tabular}{|c|c|c|c|c|}
    \hline
    \begin{tabular}[c]{@{}c@{}}\textbf{Hardware} \\ \textbf{Config}\end{tabular} & \textbf{Operator} & \textbf{Baseline} & \begin{tabular}[c]{@{}c@{}}\textbf{Cases with} \\ \textbf{Speedup > 1 (\%)}\end{tabular} & \begin{tabular}[c]{@{}c@{}}\textbf{Average} \\ \textbf{Speedup}\end{tabular} \\ 
    \hline
    \multirow{4}{*}{CPU} & \multirow{2}{*}{GEMM} & oneDNN & $77.3\%$ & $1.82\times$ \\ 
    \cline{3-5} 
    & & ONNX Runtime & $91.5\% $ & $4.38\times$ \\ 
    \cline{2-5} 
    & \multirow{2}{*}{Conv.} & oneDNN & $85.8\%$ & $2.09\times$  \\ 
    \cline{3-5} 
    & & ONNX Runtime & $99.1\%$ & $5.37\times$ \\ 
    \hline
    \multirow{4}{*}{\begin{tabular}[c]{@{}c@{}}GPU \\ (Tensor Core Enabled)\end{tabular}} & \multirow{2}{*}{GEMM} & cuBLAS & $83.7\%$ & $1.43\times$ \\ 
    \cline{3-5} 
    & & CUTLASS & $94.2\%$ & $2.62\times$ \\ 
    \cline{2-5} 
    & \multirow{2}{*}{Conv.} & cuDNN & $89.9\%$ & $2.32\times$ \\ 
    \cline{3-5} 
    & & CUTLASS & $80.5\%$ & $1.70\times$ \\ 
    \hline
    \multirow{6}{*}{\begin{tabular}[c]{@{}c@{}}GPU \\ (Cuda Core Only)\end{tabular}} & \multirow{3}{*}{GEMM} & cuBLAS & $78.3\%$ & $1.63\times$ \\

    \cline{3-5} 
    & & CUTLASS & $99.8\%$ & $7.65\times$ \\ 
    \cline{3-5} 
    & & DietCode & $94.1\%$ & $2.67\times$ \\ 
    \cline{2-5} 
    & \multirow{3}{*}{Conv.} & cuDNN & $91.1\%$ & $1.53\times$ \\

    \cline{3-5} 
    & & CUTLASS & $87.8\%$ & $2.88\times$ \\ 
    \cline{3-5} 
    & & DietCode & $92.5\%$ & $3.39\times$ \\ 
    
    \hline
    
    \end{tabular}
    
    }
    
    \end{table}

\subsection{Dynamic-Shape Tensor Program}

In this subsection, we present the evaluation of single dynamic-shape tensor program, specifically assessing GEMM and Convolution operators on CPU and GPU platforms. 
It is notable that DietCode is limited to GPU CUDA Cores and requires pre-determination of dynamic shape samples. 
We leverage the parameters from \autoref{tbl:benchmark_gemm} and \autoref{tbl:benchmark_conv} as sample sets for DietCode's offline compilation process.
Importantly, the latency measurement for \proj{} encompasses both the operator execution time on the hardware platforms and the runtime overhead from \proj{}'s cost model.

The evaluation results, shown in \autoref{fig:op_benchmark_rst}, demonstrate \proj{}'s performance across various configurations.
The x-axis outlines the number of floating-point operations (FLOPs) in the workloads, including all GEMM test cases from \autoref{tbl:benchmark_gemm} and convolution from \autoref{tbl:benchmark_conv}, while the y-axis represents the speedups.
\proj{} consistently achieves a generalized performance speedup, demonstrating improvements across different hardware setups, operators, and tensor shapes.
To further quantify the effectiveness of \proj{}, we emphasize two metrics: the percentage of cases where \proj{} shows performance improvement (defined as cases where the speedup is greater than one) and the average speedup across all cases. 
\autoref{tbl:op_eval_rst} presents these detailed results.
Overall, this comprehensive evaluation confirms that \proj{} provides robust and efficient acceleration results.

\begin{table}[b]
    \centering
    \small

    \caption{\small Speedups of \proj{} over DietCode for GEMM on GPU across different runtime ranges of M dimension.}
    \label{tbl:dietcode_eval}

    \begin{tabular}{|llll|}
    \hline
    \multicolumn{4}{|c|}{96   Test Cases: M $\in$ {[}1,384{]}, N = 768, K = 2304}                                                \\ \hline
    \multicolumn{1}{|l|}{Input Range for M} & \multicolumn{1}{l|}{{[}0, 128)} & \multicolumn{1}{l|}{{[}128, 256)} & {[}256, 384) \\ \hline
    \multicolumn{1}{|l|}{Avg. Speedups}    & \multicolumn{1}{l|}{2.8x}       & \multicolumn{1}{l|}{1.4x}         & 2.1x         \\ \hline
    \end{tabular}

    \end{table}

Additionally, we explore the impact of DietCode's reliance on sample-specific performance. 
As shown in \autoref{tbl:dietcode_eval}, we configure the M dimension dynamically in DietCode, sampling and compiling it within the range [128, 256). 
The results show a performance decline when deviating from this range, highlighting DietCode's limited flexibility.

\subsection{Dynamic-Shape Network}

\begin{figure}[t]
    \centering

    \includegraphics[width=\linewidth]{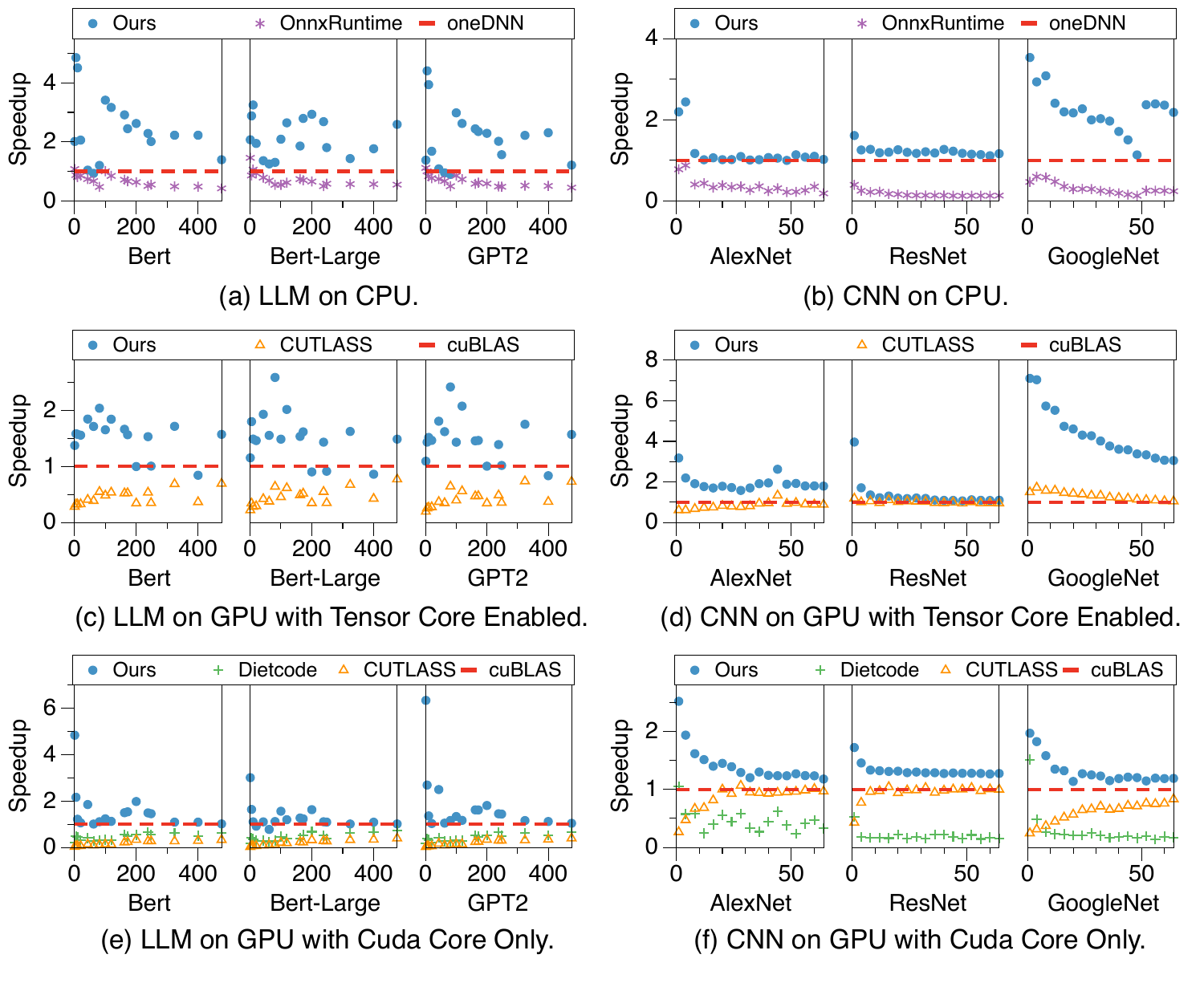}

    \caption{\small Performance results at model-level.
    The x-axis represents sequence length for LLM and batch size for CNN, and the y-axis quantifies the speedups achieved relative to the baselines. }
    
    \label{fig:eval:e2e}
    
\end{figure}

\autoref{fig:eval:e2e} presents a comprehensive evaluation of the end-to-end performance of classic language models and CNN models.
We compare \proj{} against existing state-of-the-art solutions, using oneDNN and cuBLAS/cuDNN as the normalized baselines for CPU and GPU evaluations, respectively.

\proj{} demonstrates significant performance improvements across various tasks.
\proj{} achieves notable average speedups of $2.91\times$ for BERT, $2.63\times$ for BERT-Large, and $2.94\times$ for GPT-2, across different baselines and hardware configurations. For CNNs, \proj{} achieves remarkable average speedups of $2.01\times$ for AlexNet, $2.13\times$ for ResNet, and $3.24\times$ for GoogleNet.
Furthermore, \proj{} demonstrates considerable performance improvements over multiple existing solutions. 
Specifically, \proj{} achieves average $1.73\times$, $4.26\times$, $1.43\times$, $1.71\times$, $3.32\times$ and $4.13\times$ over oneDNN, ONNX Runtime, cuBLAS, cuDNN, CUTLASS, DietCode, respectively.

The evaluation results underscore that the performance of \proj{} varies with different hardware and models, reflecting the distinct execution characteristics inherent to each environment. Notably, \proj{} consistently outperforms the baseline across a wide range of input shapes and model types, demonstrating its exceptional adaptability and efficiency in real-world dynamic-shape DNN computation.

\subsection{Additional Analysis}
\label{subsec:addition_analysis}

\paragraph{Offline Overhead Analysis.}
We first analyze the offline period's overhead. For diverse tensor shapes, \proj{} requires only a single compilation process, substantially reducing compilation overhead. For instance, in the GEMM evaluation across three computation modes (CPU, GPU with \textit{Tensor Core Enabled} mode, and GPU with \textit{Cuda Core Only}), \proj{} employs the candidates generation algorithm (\S\ref{subsection:candidate_generate}) to yield 17731, 392, and 2332 distinct candidates. The respective time overheads are 29.3s, 92.2s, and 529.6s. 
In contrast, DietCode incurs a tuning duration of 25 hours in the \textit{Cuda Core Only} mode, using configurations in \autoref{tbl:benchmark_gemm} as the sample set. \proj{} thus achieves $174\times$ enhancement in compilation efficiency compared to DietCode. 
This reduced overhead is attributable to the efficient hardware-aware pruning of candidates and the utilization of analytical cost analysis, thereby obviating the need for extensive traditional profiling.

\begin{figure}[t]
    \centering
    \includegraphics[width=0.98\linewidth]{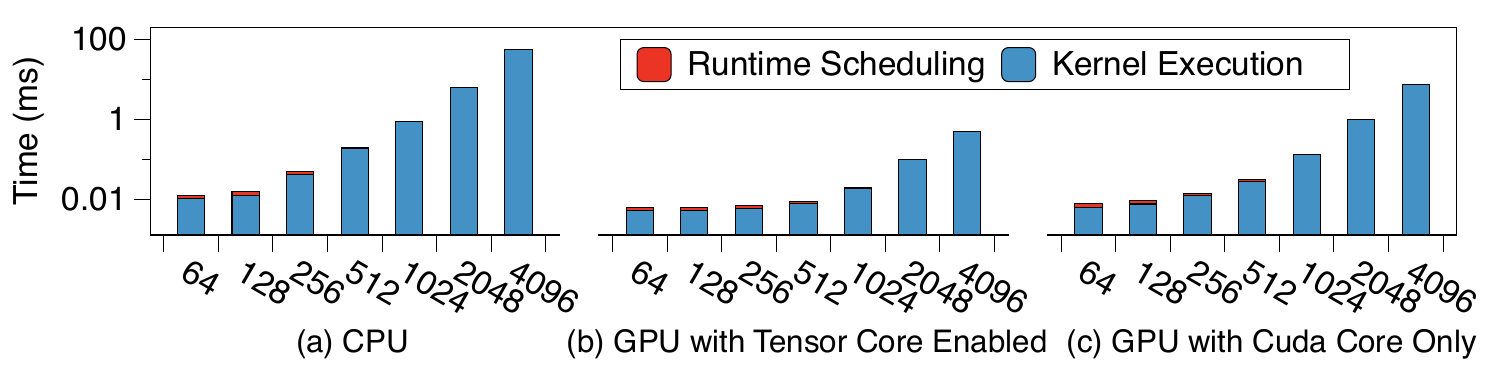}

    \caption{\small Runtime overhead breakdown of \proj{} in GEMM. The x-axis represents various M/N/K parameters, and the y-axis represents execution time.}
    \label{fig:eval:cost_model}

\end{figure}

\paragraph{Runtime Overhead Analysis.}
The primary source of runtime overhead in \proj{} is the increased computational requirements of the cost model. As depicted in \autoref{fig:eval:cost_model}, the breakdown of \proj{}’s execution on GPUs is presented, highlighting both the runtime scheduling costs and the execution durations of the final tensor programs for different shapes. The GEMM test is conducted with M/N/K values ranging from 64 to 4096. 
Notably, this runtime overhead impact is remarkably slight across various hardware platforms, demonstrating the significant runtime efficiency of \proj{}.

\begin{figure}[t]
    \centering
    \includegraphics[width=0.98\linewidth]{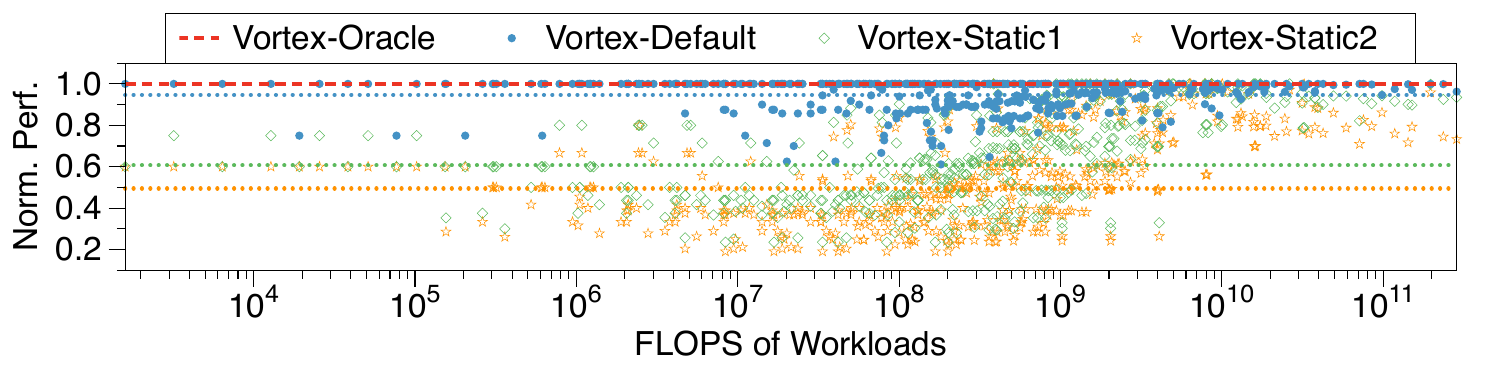}

    \caption{\small Performance comparison in GPU \textit{Tensor Core Enabled} mode. The x-axis represents the FLOPS of workloads, and the y-axis shows performance normalized to \textit{Vortex-Oracle}. Dashed lines show the normalized average performance across test cases. }
    \label{fig:eval:hw_hierarchy}

\end{figure}

\paragraph{Hierarchical Kernel Construction Evaluation.}
To validate the effectiveness of \proj{}'s dynamic and hierarchical kernel construction methodology, we assessed its default configuration against three variants:
\textit{Vortex-Oracle}, which utilizes \proj{} as a static-shape compiler with a profiling-based analyzer for all layers in every test case from \autoref{tbl:benchmark_gemm};
\textit{Vortex-Static1}, which maintains dynamic strategies at the L1 layer and adopts a static configuration for the L0 layer, selecting the most frequently optimal strategy;
\textit{Vortex-Static2}, which disables dynamic strategy selection at both the L0 and L1 layers, applying the same fixed strategy as \textit{Vortex-Static1}.
As shown in \autoref{fig:eval:hw_hierarchy}, \proj{} achieves an average of 94.7\% of the performance of \textit{Vortex-Oracle}. Meanwhile, \textit{Vortex-Static1} and \textit{Vortex-Static2} achieve 60.7\% and 49.5\% of \textit{Vortex-Oracle}'s performance, respectively. 
These experimental results underscore the effectiveness of \proj{}'s dynamic code generation and emphasize the importance of maintaining dynamic strategies across varying hardware hierarchies.

\begin{table}[b]
    \centering

    \caption{\small Comparison of \proj{}'s default configuration with the modified analyzer setup.
    'E' denotes layers using the empirical method, while unlabeled layers use the analytical method.
    }
    \label{tbl:eval:analyzer}

    \resizebox{0.49\textwidth}{!}{
    
    \begin{tabular}{|c|c|c|c|}
    \hline
    HW.                                                                                       & Analyzer Config.                                                           & \begin{tabular}[c]{@{}c@{}}Offline\\      Overhead\end{tabular} & \begin{tabular}[c]{@{}c@{}}Execution\\      Performance\end{tabular} \\ \hline
    \multirow{2}{*}{CPU} & Default (E: L0) & 29.3 sec & $1\times$ \\ \cline{2-4} 
     & Changed (E: L0, L1) & 33.0 hour & $1.04\times$  \\ \hline 
    \multirow{2}{*}{\begin{tabular}[c]{@{}c@{}}GPU\\ (\textit{Tensor Core Enabled})\end{tabular}} & Default (E: L0, L1) & 92.2 sec & $1\times$ \\ \cline{2-4} 
     & Changed (E: L0) & 19.4 sec & $0.84\times$ \\ \hline 
     \multirow{2}{*}{\begin{tabular}[c]{@{}c@{}}GPU\\ (\textit{Cuda Core Only})\end{tabular}} & Default (E: L0, L1) & 529.6 sec & $1\times$ \\ \cline{2-4} 
     & Changed (E: L0) & 39.1 sec & $0.63\times$ \\ \hline
    \end{tabular}
    }

    \end{table}

\paragraph{Hybrid Analyzer Evaluation.}
We conduct a study to assess the effectiveness of the hybrid analyzer in \proj{}. 
We compare the default \proj{} with a modified configuration as detailed in \autoref{tbl:eval:analyzer}. 
We collect the data on offline compilation overhead, and the average runtime performance for all cases in \autoref{tbl:benchmark_gemm}.
The experimental results reveal a significant and sharp increase in the CPU's offline overhead when the configuration is modified, resulting in only marginal performance gains.
Conversely, this modification leads to a significant reduction in GPU performance.
These findings substantiate and underscore the rationale behind choosing the default configuration as the preferred setup for \proj{}.

\begin{figure}[t]
    \centering
    \includegraphics[width=0.98\linewidth]{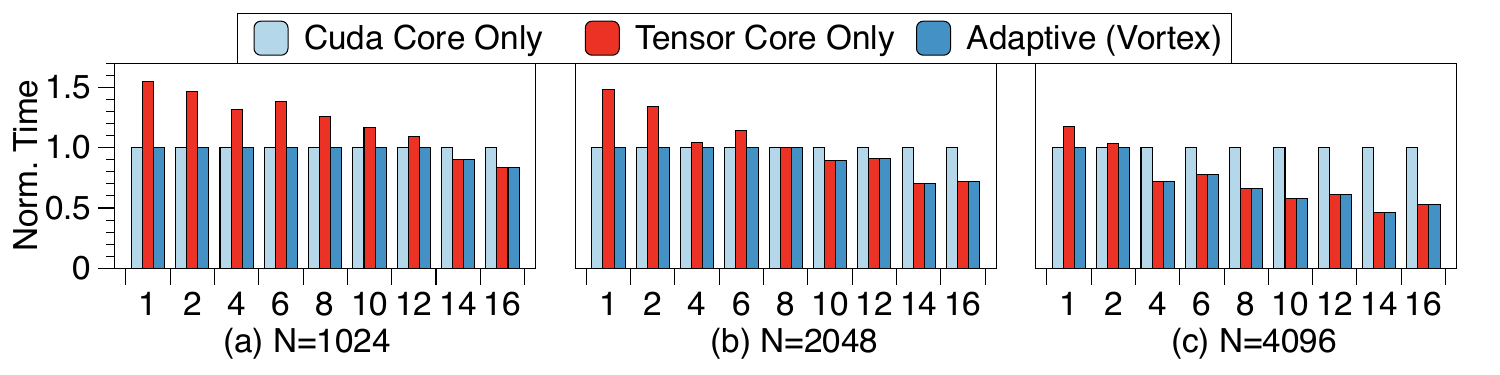}

    \caption{\small Performance comparison on GPU across \textit{Cuda Core Only}, \textit{Tensor Core Only}, and \textit{Adaptive} modes for GEMM. The x-axis represents the M value, and the y-axis shows the execution time normalized to the CUDA Core Only mode.}
    \label{fig:adaptive_hardware}

\end{figure}

\paragraph{Dynamic Hardware Adaptation.}
We investigate the dynamic hardware adaptability of \proj{} on GPUs using FP16 for the GEMM operator.
We test N values of 1024, 2048, and 4096, with K fixed at 1024 and M dynamically adjusted from 1 to 16, across three settings: \textit{Cuda Core Only}, \textit{Tensor Core Only}, and the default \textit{Adaptive}.
The results, presented in \autoref{fig:adaptive_hardware}, reveal dynamic hardware utilization opportunities and clarify hardware selection criteria for specific scenarios. 
\proj{} effectively utilizes optimal hardware, achieving performance gains of up to 48\% and 54\% over fixed CUDA and Tensor Core settings, respectively, demonstrating the effectiveness of its hardware-adaptive scheduler.

\section{Related Work} 
For dynamic-shape tensor programs, vendor-provided libraries, such as cuBLAS~\cite{cublas}, cuDNN~\cite{cudnn}, MKL-DNN~\cite{mkl} and CUTLASS~\cite{cutlass} are extensively utilized in prevalent frameworks, facilitating high-performance tensor operations across diverse hardware platforms. These libraries, tailored to specific target hardware, require substantial engineering efforts. \proj{}, by introducing a novel unified recursive abstraction, significantly reduces development costs and unifies optimization strategies across different hardware platforms.

Additionally, compilation optimization is a crucial solution for dynamic-shape tensor programs.
Existing methods, such as DietCode~\cite{dietcode}, Nimble~\cite{nimble}, and DISC~\cite{bladedisc}, predominantly rely on sample-based compilation approaches.
However, these methods overlook hardware-aware optimization opportunities. Our work, distinguishing itself in this area, uses hardware information as a fundamental element to construct a novel sample-free compilation workflow, thus supporting diverse high-performance scenarios.

For optimizing tensor programs with static shapes, various tensor compilers such as AutoTVM~\cite{autotvm}, FlexTensor~\cite{flextensor}, Ansor~\cite{ansor}, and TensorIR~\cite{tensorir} have been proposed. 
However, these methods are associated with significant compile time overheads. Although efforts like Roller~\cite{roller} have attempted to optimize the compilation time for static-shape compilers, the time required is still considerably longer than the execution overhead, making them impractical for the online demands of dynamic-shape tensor programs.

Graph-level optimization is another important component for end-to-end DNN optimizations. 
DNNFusion~\cite{DNNFusion}, Rammer~\cite{Rammer}, Chimera~\cite{chimera}, and AStitch~\cite{astitch} focus on fusion optimizations in DNNs.
TASO~\cite{taso}, Unity~\cite{unity}, XLA~\cite{XLA}, JAX~\cite{jax}, TorchDynamo~\cite{TorchDynamo}, TenSAT~\cite{tensat} explore graph rewriting opportunities.
In this paper, our proposed \proj{} focuses on operator-level for dynamic-shape tensor programs, which is orthogonal to these works.
Meanwhile, \proj{} is designed without inherent limitations that hinder integration with current graph-level compilers.
We look forward to exploring this area as a part of our future research efforts.

\section{Conclusion}

In this work, we propose \proj{}, an hardware-driven and sample-free dynamic-shape tensor program compiler.
\proj{} leverages bidirectional compilation techniques to deliver universally high-performance support with minimal system overhead. 
Experimental results demonstrate that \proj{} achieves average speedups of $2.53\times$ and $3.01\times$ over vendor-provided libraries and existing dynamic-shape compiler, respectively. 
These results highlight the effectiveness of \proj{} and its potential as
a standard methodology for enhancing dynamic-shape tensor program optimization.

\bibliographystyle{plain}
\bibliography{ref}

\end{document}